\documentclass[twocolumn,preprintnumbers,floats,prd,amssymb,floatfix,nofootinbib,balancelastpage,superscriptaddress,amsmath]{revtex4-1}

\allowdisplaybreaks[4]
\usepackage[utf8]{inputenc}
\usepackage{amssymb}
\usepackage{amsmath}
\usepackage{amsfonts}
\usepackage{graphicx}
\usepackage{color}
\usepackage{xspace}
\usepackage{comment}
\usepackage{hyperref}
\usepackage[normalem]{ulem}
\usepackage[section]{placeins}
\usepackage{afterpage}
\usepackage{slashed}
\usepackage{enumerate}
\usepackage{enumitem}
\usepackage{caption}
\usepackage{subfigure}
\usepackage{float}
\usepackage{ulem}
\usepackage{verbatim}
\usepackage{multirow,rotating}
\usepackage[dvipsnames]{xcolor}
\usepackage{braket}

\definecolor{dcolour}{rgb}{.5, .5, .5}
\def\gsim{\raise0.3ex\hbox{$\;>$\kern-0.75em\raise-1.1ex\hbox{$\sim\;$}}}
\def\lsim{\raise0.3ex\hbox{$\;<$\kern-0.75em\raise-1.1ex\hbox{$\sim\;$}}}
\def\gsim{\raise0.3ex\hbox{$\;>$\kern-0.75em\raise-1.1ex\hbox{$\sim\;$}}}
\def\lsim{\raise0.3ex\hbox{$\;<$\kern-0.75em\raise-1.1ex\hbox{$\sim\;$}}}

\newcommand{\ba}[1]{\begin{eqnarray} \label{(#1)}}
\newcommand{\ea}{\end{eqnarray}}

\newcommand{\added}[1]{\textcolor{red}{#1}}

\begin{document}

\title{Investigate the strong coupling of \texorpdfstring{$g_{X J/\psi\phi}$}{Lg} in \texorpdfstring{$X(4500) \to J/\psi \phi$}{Lg} by using the three-point sum rules and the light-cone sum rules}

\author{\textsc{Yiling Xie}}
\email{xieyl_9@mail.dlut.edu.cn}
\affiliation{Institute of Theoretical Physics, School of Physics, Dalian University of Technology, \\ 
No.2 Linggong Road, Dalian, Liaoning, 116024, People’s Republic of China}
\author{\textsc{Hao Sun}}
\email{haosun@dlut.edu.cn}
\affiliation{Institute of Theoretical Physics, School of Physics, Dalian University of Technology, \\ 
No.2 Linggong Road, Dalian, Liaoning, 116024, People’s Republic of China}


\begin{abstract}
We assign $X(4500)$ as a D-wave tetraquark state and study the decay of $X(4500)$ $\to$ $J/\psi \phi$. The mass and the decay constant of $X(4500)$ are calculated by using the SVZ sum rules.
For the decay width of $X(4500)$ $\to$ $J/\psi \phi$, we present the calculation
within the framework of both the three-point sum rules and the light-cone sum rules. 
The strong coupling $g_{X J/\psi \phi}$ is obtained by considering the soft-meson approximation when we use the light-cone sum rules calculation.
Both calculations show that the decay of $X(4500)$ $\to$ $J/\psi\phi$ close to the total width of $X(4500)$ if we assign $X$(4500) as a D-wave tetraquark.
We also consider the open-charm decay channels like $X(4500)\to D_sD_s$, $X(4500)\to D_s^*D_s$, and $X(4500)\to  D_s^*D_s^*$ , their widths are small when compared to the width of $X(4500)$, suggesting that the hidden-charm decay channels $X(4500)$ $\to$ $J/\psi\phi$ are predominant when compared with the total width of $X(4500)$. A more rational conclusion can be obtained only if the complete open-charm decay channel is considered. In the future, experiments will be more helpful in determining whether or not this structure of $X(4500)$ is appropriate.

\end{abstract}
\vskip10mm
\maketitle
\flushbottom
\tableofcontents
%
%

\section{Introduction}
\label{sec:intro}

In 2016, the LHCb collaboration analyzed the decay of $B^{+}$ $\to J/\psi\phi K^+$ in the $pp$ collision~\cite{LHCb:2016axx,LHCb:2016nsl},
and found that four resonance-like peaks appeared in the $J/\psi\phi$ invariant mass spectrum. 
Two of them are named $X(4140)$ and $X(4274)$, and their spin parity numbers are $J^{PC}=1^{++}$. 
The $J^{PC}$ of the other two, with the names of $X(4500)$ and $X(4700)$ respectively, are $J^{PC}=0^{++}$. 

It is necessary to investigate $X(4140)$ and $X(4274)$, $X(4700)$ and $X(4500)$ because they may make up a significant fraction of exotic mesons. Due to the fact that the mass spectrum of $X(4140)$ and $X(4274)$, and the hidden-flavor decay widths of $X(4700)$ have already been calculated with sum rules \cite{Agaev:2017foq,Xie:2022ilz}, in this paper, we focus on $X(4500)$, whose mass and width are $\text{M}=4474\pm 3\pm 3\text{MeV}$ and $\Gamma=77\pm6^{+10}_{-8}\text{MeV}$ respectively~\cite{LHCb:2021uow}.

It is now the structure of X(4500) has not yet been fully established from a variety of studies, that many more works are in progress on this attractive issue.

First, it is interesting to ascertain whether the re-scattering effects may contribute to the resonance peaks in the $J/\psi \phi$ mass spectrum or not\cite{Luo:2022xjx}.
Through investigating this effects or threshold cusps in the process $B^+\to J/\psi\phi K^+$, 
the authors in Ref.\cite{Liu:2016onn,Ge:2021sdq} found that, for $X$(4700), 
it can be simulated by the $\psi^\prime\phi$ re-scattering via the $\psi^\prime K_1$ loops, 
while for $X(4500)$, it is difficult to attribute it to such effects. 
It may correspond to genuine resonances instead.

Since $X$(4500) appears in the $J/\psi\phi$ invariant mass spectrum, we could speculate that it can be a charmonium. 
Whether $X(4500)$ is assigned as a $4^3P_0$ state in the constituent quark model~\cite{Fernandez:2016bqr,Ortega:2016hde}
 or assigned as a $\chi_{c0}(4P)$ state in the linear potential model\cite{Gui:2018rvv}, 
its predicted mass and width are consistent within the experimental uncertainties.
The theoretical calculation shows that $4^3P_0$ will dominantly decay into charm mesons like $DD_1$, $DD_1^{\prime}$, $D^*D^\prime$, and $D^*D^*_2$, 
and that $\chi_{c0}(4P)$ will dominantly decay into $DD_1$, $DD_1^\prime$, $D^*D_0$, $D^*D_2$. 
However, in experiemnt, no relevant signal of $X(4500)$ appears in the $B\to$ $D^{(*)}_{(s)}\bar{D}^*_{(s)}K$ decays~\cite{Olsen:2017bmm}.
Besides, since $\chi_{cJ}(nP)$ $\to$ $J/\psi\omega$ is similar to $X(4500)\to J/\psi\phi$,
we can surmise the $J/\psi\omega$ mass spectrum in $B^{+}$ $\to J/\psi\omega K^+$ decay should appear some structures resembling the $J/\psi\phi$ mass peaks. 
But this was rejected by experiment~\cite{BaBar:2012nxg,Belle:2009and}, which contradicts a charmonium interpretation for the $X(4500)$.
In addition, in the screened potential model \cite{Gui:2018rvv}, 
$\chi_{c0}(5P)$ is a good candidate for $X$(4500), since its predicted mass is $4537$ MeV. 
Whereas, the total width of $\chi_{c0}(5P)$ is about $15$ MeV, 4 times smaller than that of $X$(4500) observed in the experiment.

Although $X(4500)$ is inconsistent with pure charmonium, it could be $J/\psi-\phi$ or $\Psi(2S)-\phi$ bound state since it appears in the spectrum of $J/\psi\phi$.
Unfortunately, lattice QCD exhibited weak attraction of $J/\psi-\phi$ at low energies~\cite{Ozaki:2012ce},
and $\Psi(2S)-\phi$ bound state description has already been occupied by $X(4274)$~\cite{Panteleeva:2018ijz}.
In some specific parameter space, $X(4500)$ might be assigned as other hadrocharmonium states but require different binding mechanisms~\cite{Panteleeva:2018ijz}.
It can also be expanded as hybrid state which is a charmonium state that incorporates excited gluon fields.
For example, the mass of $X(4500)$ is compatable with that of $0^{++}$ hybrid $1p_0(H_3)$ state in the NRQCD results~\cite{Oncala:2017hop}. 
But it's tricky to expalin $X(4500)$ observed in $J/\psi\phi$ channel, since $1p_0(H_3)$ only contains small pieces of charmonium.
  
In addition to these charmonium-related explanations, 
some theoretical studies are oriented toward the $cs\bar{c}\bar{s}$ tetraquark state.

Reported by Maiani et al.\cite{Maiani:2016wlq}, $X(4500)$ can be assigned as radial excitations of $X(4140)$ while $X(4140)$ was arranged as a $cs\bar{c}\bar{s}$ ground state.
Similarly, in Ref.\cite{Lu:2016cwr}, it was regarded as a tetraquark that incorporates one 2S scalar diquark and one scalar antidiquark within the relativized quark model.
In other models like the diquark-antidiquark model~\cite{Zhu:2016arf}, $X(4500)$ was explained as a radial excitation of $J^P=0^+$ tetraquark 
with quark content $\frac{1}{\sqrt{3}}(u\bar{u}+d\bar{d}+s\bar{s})c\bar{c}$, 
and in the chiral quark model~\cite{Yang:2019dxd}, it was explained as a 2S radial excited compact tetraquark state with $J^{PC}=0^{++}$.
Besides, as illustrated in the color flux-tube model with a multibody confinement potential~\cite{Deng:2017xlb},
$X(4500)$ can be probably considered as a $0^{++}$ compact tetraquark states $cs\bar{c}\bar{s}$ although its mass is slightly higher than $4500$ MeV.
In the framework of the quark delocalization color screening model~\cite{Liu:2021xje}, 
$X(4500)$ has been assigned as a $IJ^P = 00^+$ compact tetraquark resonance state,
or a radial excitions of S-wave scalar diquark-antidiquark bound states with quantum number $0^{++}$ in the diquark model~\cite{Anwar:2018sol}.
Moreover, in the sum rules approach~\cite{Reinders:1984sr}, 
it can also be considered as a first radial excited state~\cite{Wang:2016gxp} or a D-wave $cs\bar{c}\bar{s}$ tetraquark states~\cite{Chen:2016oma}.

As we mentioned, with the method of SVZ sum rules, $X$(4500) has been investigated in Ref.\cite{Chen:2016oma} in which its mass is predicted.
Because of the deficiency of hidden-charm decay width and open-charm decays width, we consider the hidden-charm decay and open-charm decay channel in this article.
Following the assumption in the Ref.\cite{Chen:2016oma},
we consider $X$(4500) a D-wave tetraquark state in our study.
 When we assign $X(4500)$ as a D-wave tetraquark, $X(4500)$ can possibly decay into $D_sD_s$, $D_s^*D_s$, and $D_s^*D_s^*$. Furthermore, by performing the Fierz and color rearrangements on the currents and changing it to mesonic-mesonic structures, the decay channels associated with $X(4500)$ can be obtained\cite{Chen:2016qju,Chen:2006hy,Chen:2007xr}. We can list the final state as follows:
to be S-wave $D^*_s D^*_{s1}(2860)$, $D^*_s D^*_{s3}(2860)$, P-wave $D^*_s D^*_{s0}$, $D^*_s D_{s1}$, $D^*_s D^*_{s2}$ , and D-wave $D_s^*D_s^*$ and etc. Since only the D  meson light cone distribution amplitude (LCDA) is well defined in Refs \cite{Zuo:2006re,Li:2008ts}, there is no systematic development of the D family meson LCDAs in the literature, so we  deal with some open-charm decays like $D_sD_s$, $D_s^*D_s$, and $D_s^*D_s^*$ with three-point sum rule. Some other decays involving $D_{sJ}^*,J=0,1,2$ are much more troublesome because $D_{sJ}^*,J=0,1,2$ structures are complicated and unstable. Therefore, the only open-charm decay channels we consider in this paper are the $X(4500)\to D_sD_s$, $X(4500)\to D_s^*D_s$, and $X(4500)\to  D_s^*D_s^*$ .
We calculate the mass of $X(4500)$ by conducting the dimension regulation in SVZ sum rules at first, and obtain the results comparable with those in PDG \cite{ParticleDataGroup:2020ssz}.
We then evaluate its decay constant which is used in the numerical calculation of the strong coupling $g_{X J/\psi\phi}$.
Typically, we calculate $g_{X J/\psi\phi}$ through both the three-point sum rules and the light-cone sum rules (LCSR) methods simultaneously.
According to our calculations, different methods have congruent decay widths of $X\text{(4500)}\to$ $J/\psi\phi$.
We compared the results with the total widths of $X$(4500) with the experiment and demonstrate some valuable discussions. 
The results are instructive for future experiments to further determine the structure of $X(4500)$.
To previous works on $X$(4500) \cite{Chen:2016oma}, our study can be perceived as a supplement.

The structure of the paper is arranged as follows.
In Section.\ref{II}, the strong coupling $g_{XJ/\psi \phi}$ is obtained
by using both the three-point sum rules and the light-cone sum rules. 
Through the two-point SVZ sum rules, we study the mass and decay constant of $X$(4500). 
The numerical results and discussions are provided in Section.\ref{III}. 
We come to the summary in Section.\ref{sec:summary}.

\section{Calculation Framework}
\label{II}

\subsection{ The mass and the decay constant of \texorpdfstring{$X(4500)$}{Lg} }

In this section, we calculate the mass and the decay constant of $X(4500)$ to extract their values. 
The two-point function provides the foundation for the sum rule calculation of the mass:
\begin{equation}\label{corm}
    \begin{aligned}
    \Pi^{\text{SVZ}}(p)=i\int \mbox{d}^4xe^{ipx}
    \braket{0|T\{J^{X}(x)J^{X\dagger}(0)\}|0},
    \end{aligned}
\end{equation}
The interpolating currents of $X(4500)$ is given by\cite{Chen:2016oma}:
\begin{eqnarray}\label{Xcurrents}\nonumber
        &J^{X}(x)&=c_k^T(x)C\gamma_{\mu 1}[\overrightarrow{D}_{\mu 3}\overrightarrow{D}_{\mu 4}s_l(x)]\\ \nonumber
        &\quad\quad\quad&\quad  (\bar{c}_k(x)\gamma_{\mu 2}Cs_l^T(x)-\bar{c}_l(x)\gamma_{\mu 2}C\bar{s}_k^T(x))\\ \nonumber
        &\quad\quad\quad  &\times(g^{\mu 1\mu 3}g^{\mu 2\mu 4}+g^{\mu 1\mu 4}g^{\mu 2\mu 3}-g^{\mu 1\mu 2}g^{\mu 3\mu 4}/2),
\end{eqnarray}
where $\overrightarrow{D}_\mu=\overrightarrow{\partial}_\mu +igT^jA_\mu^j$.
The subscripts $s, c$ refer to the strange and charm quarks, and the subscripts $j$, $k$, $l$ to the color indices, and $C$ for the charge conjugation matrix.

A phenomenological expression of the correlation function can be given by considering the complete set of hadronic states:
\begin{equation}\label{pho}
    \begin{aligned}
        &\Pi^{\text{SVZ,phen}}(p)=\\
        &\frac{\braket{0|J^{X}|X(p)}
        \braket{X(p)|J^{X\dagger}|0}}{m_X^2-p^2}
        +\int_{s^\prime}^\infty d \hat{s}\frac{\tilde{\rho}^{\text{SVZ,phen}}(\hat{s})}{\hat{s}-p^2},
    \end{aligned}
\end{equation}
Here the higher resonances and continuum states are referred to as 
$\tilde{\rho}^{\text{SVZ,phen}}$. 
Since the subtraction terms will vanish after Borel transformation, they are not shown here.

Next, 
by replacing all matrix elements with
\begin{eqnarray}\label{decay-cons}
        &&\braket{0|J^X|X(p^\prime)}=m_Xf_X\\
\end{eqnarray}
and performing the Borel transformation, the Eq.\eqref{pho} can be written as
\begin{equation}
    \begin{aligned}
        &\Pi^{\text{SVZ,phen}}(M^2)=\\
        &(m_X)^2(f_X)^2e^{-(m_X)^2/M^2}+\int_{s^\prime}^\infty d\hat{s}\tilde{\rho}^{\text{SVZ,phen}}(\hat{s})e^{-\hat{s}/M^2}.
    \end{aligned}
\end{equation}

Now we'll establish the correlation function on the OPE side, where the non-vanishing vacuum expectation of quark and gloun condensate
such as $\langle \bar{q}q \rangle$,$\langle \bar{q}g_s\sigma Gq \rangle$ are introduced.
First, we express the correlation function with interpolating currents in Eq.\eqref{Xcurrents}, perform the Wick Theorem, then the correlation function can be recast to be
\begin{equation}\label{propagator2}
    \begin{aligned}
        &\Pi^{\text{SVZ,OPE}}(p)=i\int \mbox{d}^4xe^{ipx}\\
        &\text{Tr}[\gamma_{\mu 1}(\overrightarrow{D}_{\mu 3}\overrightarrow{D}_{\mu 4}S^{kk^\prime}_s(x-y)\overleftarrow{D}_{\nu 3}\overleftarrow{D}_{\nu 4})\gamma_{\nu 1}\gamma_5\tilde{S}_c^{jj^\prime}(-x)]|_{y=0}\\
        &(\text{Tr}[\gamma_{\mu 2}\tilde{S}^{kk^\prime}_s(-x)\gamma_{\nu 2}S_c^{jj^\prime}(-x)]\\
        &-\text{Tr}[\gamma_{\mu 2}\tilde{S}^{kj^\prime}_s(-x)\gamma_{\nu 2}S_c^{jk^\prime}(-x)]\\
        &-\text{Tr}[\gamma_{\mu 2}\tilde{S}^{jk^\prime}_s(-x)\gamma_{\nu 2}S_c^{kj^\prime}(-x)]\\
        &+\text{Tr}[\gamma_{\mu 2}\tilde{S}^{jj^\prime}_s(-x)\gamma_{\nu 2}S_c^{jj^\prime}(-x)])\\
        \times&(g^{\mu 1\mu 3}g^{\mu 2\mu 4}+g^{\mu 1\mu 4}g^{\mu 2\mu 3}-g^{\mu 1\mu 2}g^{\mu 3\mu 4}/2)\\
        \times&(g^{\nu 1\nu 3}g^{\nu 2\nu 4}+g^{\nu 1\nu 4}g^{\nu 2\nu 3}-g^{\nu 1\nu 2}g^{\nu 3\nu 4}/2).
    \end{aligned}
\end{equation}
In this case, one must deal with divergences in the double integrals like:
\begin{equation}
    \begin{aligned}
        \int \frac{d^4x}{x^{2n}}\int\int\frac{d^4k_1d^4k_2e^{ipx-ik_1x-ik_2x}}{(k_1^2-m_c^2)(k_2^2-m_c^2)}.
    \end{aligned}
\end{equation}
what we have to do is transform the coordinate to the momentum space in D-dimensio by the Fourier transformation\cite{Agaev:2016dev}
\begin{equation}
    \begin{aligned}
        \begin{aligned}
            &\frac{1}{\left(x^{2}\right)^{n}}=\int \frac{d^{D} p}{(2 \pi)^{D}} e^{-i p \cdot x} i(-1)^{n+1} 2^{D-2 n} \pi^{D / 2} \\
                &\times \frac{\Gamma(D/ 2-n)}{\Gamma(n)}\left(-\frac{1}{p^{2}}\right)^{D / 2-n}
\end{aligned}
    \end{aligned}
\end{equation}
The results combine the with remains part dimensionally regularize at
D = 4 \cite{Matheus:2006xi}. Then we can extract the spectral densitythe form imaginary part of results.

The OPE side of the correlation function also can be written as:
 \begin{equation}\label{OPE1}
    \begin{aligned}
        \Pi^{\text{SVZ,OPE}}(p)=\int_{4m_c^2}^\infty\mbox{d}\hat{s}\frac{\tilde{\rho}^{\text{SVZ,OPE}}(\hat{s})}{\hat{s}-p^2},
    \end{aligned}
 \end{equation}
where $\tilde{\rho}^{\text{SVZ,OPE}}(\hat{s})$ corresponds to the spectral density.

By performing the Borel transform at both the Eq.\eqref{pho} and Eq.\eqref{OPE1}, equating the obtained expression:
\begin{equation}
    \begin{aligned}
        &\int_{4m_c^2}^\infty\mbox{d}\hat{s}\tilde{\rho}^{\text{SVZ,OPE}}(\hat{s})e^{-\hat{s}/M^2}=\\
        &(m_X)^2(f_X)^2e^{-(m_X)^2/M^2}+\int_{s_0}^\infty\tilde{\rho}^{\text{SVZ,OPE}}(\hat{s})e^{-\hat{s}/M^2},
    \end{aligned}
\end{equation}
and subtracting the continuum contribution, we find the expression:
\begin{equation}
    \begin{aligned}
        (m_X)^2(f_X)^2e^{-(m_X)^2/M^2}=\int_{4m_c^2}^{s_0}\mbox{d}\hat{s}\tilde{\rho}^{\text{SVZ.OPE}}(\hat{s})e^{-\hat{s}/M^2}.
    \end{aligned}
\end{equation}
Finally the mass of the state $X(4500)$ can be obtained as
\begin{equation}
    \begin{aligned}
        (m_X)^2=\frac{\int_{4m_c^2}^{s_0}\mbox{d}\hat{s}\hat{s}\tilde{\rho}^{\text{SVZ,OPE}}(\hat{s})e^{-\hat{s}/M^2}}{\int_{4m_c^2}^{s_0}\mbox{d}\hat{s}\tilde{\rho}^{\text{SVZ,OPE}}(\hat{s})e^{-\hat{s}/M^2}}.
    \end{aligned}
\end{equation}
In Appendix \ref{appendix:B}, we provide the spectral densities for $J^{X}$.

\subsection{The strong coupling \texorpdfstring{$g_{X D_S D_S}$, $g_{X D_S D_S^*}$, $g_{X D_S^* D_S^*}$}{Lg} in the three-point sum rules}
The QCD sum rules\cite{Reinders:1984sr} allows us to describe the strong interaction at the low energy level.
We are going to consider $X$(4500) as a D-wave tetraquark state and predict the decay width of
$X(4500)$ $\to$ $D_sD_s$, $X(4500)$ $\to$ $D_sD_s^*$  and $X(4500)$ $\to$ $D_s^*D_s^*$.
The starting point of sum rules is to write down the T-ordered product of three currents for the correlation function:
\begin{eqnarray}\label{three-pointDD}
&&\Pi^{\text{DD}}\left(p^{\prime}, p, q\right) \\ \nonumber
&=&\int d^{4} x d^{4} y e^{i p \cdot x} e^{i q \cdot y}\left\langle 0\left|T\left[J^{D_s}(x) J^{D_s\dagger}(y) J^{X \dagger}(0)\right]\right| 0\right\rangle,
\end{eqnarray}

\begin{eqnarray}\label{three-pointDDs}
&&\Pi^{\text{DDs}}\left(p^{\prime}, p, q\right) \\ \nonumber
&=&\int d^{4} x d^{4} y e^{i p \cdot x} e^{i q \cdot y}\left\langle 0\left|T\left[J^{D_s}_\mu(x) J^{D_s^*\dagger}(y) J^{X \dagger}(0)\right]\right| 0\right\rangle,
\end{eqnarray}

\begin{eqnarray}\label{three-pointDsDs}
&&\Pi_{\mu \nu}^{\text{DsDs}}\left(p^{\prime}, p, q\right) \\ \nonumber
&=&\int d^{4} x d^{4} y e^{i p \cdot x} e^{i q \cdot y}\left\langle 0\left|T\left[J^{D_s^*}_\mu(x) J^{D_s^*\dagger}_\nu(y) J^{X \dagger}(0)\right]\right| 0\right\rangle,
\end{eqnarray}

where $p$, $q$ represent the four-momentum of two final state respectively.
Momentum conservation dictates that $X(4500)$ has the four-momentum of $p^\prime=p+q$.
Meanwhile the interpolating currents for $J^{D_s}$,$J^{D_s^*}_\mu$ are given by\cite{Bracco:2006xf}:
\begin{eqnarray}\label{currents}
        J^{D_s}=i\bar{c}_a\gamma_5 s_a,\  J^{D_s^*}_{\mu}=\bar{c}_b\gamma_\mu s_b.
\end{eqnarray}
where the subscripts $a$, $b$ refer to the color indices.

Next, in an attempt to evaluate the phenomenological side of the correlation function\eqref{three-pointDD}, \eqref{three-pointDDs} and \eqref{three-pointDsDs}.
we insert intermediate states for $X(4500)$, $D_s$, $D_s^*$ into Eq\eqref{three-pointDD}, \eqref{three-pointDDs} and \eqref{three-pointDsDs} and write down the correlation function as
\begin{eqnarray}\label{Three-phenoDsDs} \nonumber
&& \Pi_{\mu \nu}^{\text{DsDs}}\left(p^{\prime}, p, q\right) \\\nonumber
&=& g_{X D_s^*D_s^*} g_{\mu^{\prime} \nu^{\prime}}\left(g_{\mu \mu^{\prime}}-\frac{p_{\mu} p_{\mu^{\prime}}}{m_{D_s^*}^{2}}\right)\left(g_{\nu \nu^{\prime}}-\frac{p_{\nu} p_{\nu^{\prime}}}{m_{X}^{2}}\right) \times \\ \nonumber
&& \frac{f_{D_s^*}^2 f_{X}}{\left(p^{\prime 2}-m_{X}^{2}+i \epsilon\right)\left(p^{2}-m_{D_s^*}^{2}+i \epsilon\right)\left(q^{2}-m_{D_s^*}^{2}+i \epsilon\right)} \\ \nonumber
&&+\ldots \\ \nonumber
&=& \frac{g_{X D_s^*D_s^*} f_{D_s^*}^2 f_{X}}{\left(p^{\prime 2}-m_{X}^{2}+i \epsilon\right)\left(p^{2}-m_{D_s^*}^{2}+i \epsilon\right)\left(q^{2}-m_{D_s^*}^{2}+i \epsilon\right)} \times \\ \nonumber
&&\left(g_{\mu \nu}-\frac{p^\prime_{\mu} p_{\nu}+p^\prime_{\mu} p^\prime_{\nu}}{m_{X}^{2}}-\frac{p_{\mu} p_{\nu}}{m_{D_s^*}^{2}}+\frac{p^\prime \cdot p\left(p_{\mu} p_{\nu}+p^\prime_{\nu} p_{\mu}\right)}{m_{X}^{2} m_{D_s^*}^{2}}\right) \\
&&+\ldots.
\end{eqnarray}
\begin{eqnarray}\label{Three-phenoDD} \nonumber
&& \Pi_{\mu \nu}^{\text{DD}}\left(p^{\prime}, p, q\right) \\\nonumber
&=& \frac{-g_{X D_sD_s} f_{D_s}^2 f_{X}m_{D_s}^4}{(m_c+m_s)^2\left(p^{\prime 2}-m_{X}^{2}+i \epsilon\right)\left(p^{2}-m_{D_s}^{2}+i \epsilon\right)\left(q^{2}-m_{D_s}^{2}+i \epsilon\right)} \\ \nonumber
&&+\ldots.
\end{eqnarray}
\begin{eqnarray}\label{Three-phenoDDs} \nonumber
&& \Pi_{\mu \nu}^{\text{DDs}}\left(p^{\prime}, p, q\right) \\\nonumber
&=& \frac{ig_{X D_s^*D_s} f_{D_s^*} f_{D_s}m_{D_s}^2m_{D_s^*} f_{X}}{(m_c+m_s)\left(p^{\prime 2}-m_{X}^{2}+i \epsilon\right)\left(p^{2}-m_{D_s^*}^{2}+i \epsilon\right)\left(q^{2}-m_{D_s}^{2}+i \epsilon\right)} \\ \nonumber
&&\times(-q^\mu+\frac{p\cdot q}{m_{D_s^*}^2}p_\mu)+\ldots.
\end{eqnarray}
Here $\dots$ denotes all the higher excited state contributions.
In order to establish Eq.\eqref{Three-phenoDsDs}, we have used the relationships\cite{Bracco:2006xf}:
\begin{eqnarray}\label{decay-cons}\nonumber
        &&\braket{0|J^X|X(p^\prime)}=m_Xf_X\\ 
        &&\braket{0|J^{D_s}|D(p)}=\frac{m_{D_s}^2}{m_c+m_s}f_{D_s}\nonumber\\
        &&\braket{0|J_{\mu}^{D_s^*}|D_s^*(p)}=m_{D_s^*}f_{D^*_s}\varepsilon_\mu,
\end{eqnarray}
in which $\epsilon_\mu$ is the polarization vector of $\phi$ and $J/\psi$.
$f_{D_s (X, D_s^*)}$ is the decay constant of $D_s (X(4500), D_s^*)$.
Besides, the coupling constant $g_{XD_s^*D_s^*}$ came from the matrix element
$\langle\phi(q)J/\psi(p)|X(p^\prime)\rangle$:
\begin{equation}\label{coupling}
    \begin{aligned}
        \langle D_s^*(q)D_s^*(p)|&X(p^\prime)\rangle\\
        =&g_{XD_s^*D_s^*}
        [(q\cdot \varepsilon^*)(p\cdot\varepsilon^\prime)-(q\cdot p)(\varepsilon^*\cdot \varepsilon^\prime)].\\
        \langle D_s(q)D_s(p)|&X(p^\prime)\rangle =g_{XD_sD_s}.\\
        \langle D_s(q)D_s^*(p)|&X(p^\prime)\rangle =g_{XD_sD_s^*}q\cdot\varepsilon.
    \end{aligned}
\end{equation}

After entering the currents into \eqref{three-pointDD}, \eqref{three-pointDDs} and \eqref{three-pointDsDs}  and applying the Wick Theorem, the OPE side of the sum rules is:
\begin{eqnarray} \nonumber
        && \Pi_{\mu \nu}^{\text{DD,OPE}}\left(p^{\prime}, p, q\right) =\int d^{4} x d^{4} y e^{i p \cdot x} e^{i q \cdot y} \\\nonumber
        && \{\text{Tr}[S^{al}_c(-x)\gamma_5 S^{ak}_s(y-z)\overleftarrow{D}_{\mu 3}\overleftarrow{D}_{\mu 4} \\ \nonumber
        &&~~~~~~\gamma_{\nu1} \tilde{S}^{bl}_c(-y)\gamma_5\tilde{S}^{bk}_s(y)\gamma_{\mu2}]\\\nonumber
        &&-\text{Tr}[S^{ak}_c(-x)\gamma_5 S^{ak}_s(y-z)\overleftarrow{D}_{\mu 3}\overleftarrow{D}_{\mu 4} \\
        &&~~~~~~\gamma_{\nu1} \tilde{S}^{bl}_c(-y)\gamma_5\tilde{S}^{bl}_s(y)\gamma_{\mu2}]\} ,
\end{eqnarray}

\begin{eqnarray} \nonumber
        && \Pi_{\mu \nu}^{\text{DsD,OPE}}\left(p^{\prime}, p, q\right) =\int d^{4} x d^{4} y e^{i p \cdot x} e^{i q \cdot y} \\\nonumber
        && \{\text{Tr}[S^{al}_c(-x)\gamma_\mu S^{ak}_s(y-z)\overleftarrow{D}_{\mu 3}\overleftarrow{D}_{\mu 4} \\ \nonumber
        &&~~~~~~\gamma_{\nu1} \tilde{S}^{bl}_c(-y)\gamma_5\tilde{S}^{bk}_s(y)\gamma_{\mu2}]\\\nonumber
        &&-\text{Tr}[S^{ak}_c(-x)\gamma_\mu S^{ak}_s(y-z)\overleftarrow{D}_{\mu 3}\overleftarrow{D}_{\mu 4} \\
        &&~~~~~~\gamma_{\nu1} \tilde{S}^{bl}_c(-y)\gamma_5\tilde{S}^{bl}_s(y)\gamma_{\mu2}]\} ,
\end{eqnarray}

\begin{eqnarray} \nonumber
        && \Pi_{\mu \nu}^{\text{DsDs,OPE}}\left(p^{\prime}, p, q\right) =\int d^{4} x d^{4} y e^{i p \cdot x} e^{i q \cdot y} \\\nonumber
        && \{\text{Tr}[S^{al}_c(-x)\gamma_\mu S^{ak}_s(y-z)\overleftarrow{D}_{\mu 3}\overleftarrow{D}_{\mu 4} \\ \nonumber
        &&~~~~~~\gamma_{\nu1} \tilde{S}^{bl}_c(-y)\gamma_\nu\tilde{S}^{bk}_s(y)\gamma_{\mu2}]\\\nonumber
        &&-\text{Tr}[S^{ak}_c(-x)\gamma_\mu S^{ak}_s(y-z)\overleftarrow{D}_{\mu 3}\overleftarrow{D}_{\mu 4} \\
        &&~~~~~~\gamma_{\nu1} \tilde{S}^{bl}_c(-y)\gamma_\nu\tilde{S}^{bl}_s(y)\gamma_{\mu2}]\} ,
\end{eqnarray}

where we have denoted
\begin{equation}
    \begin{aligned}
    \tilde{S}_q^{ab}(x)=CS_q^{ab}(x)C,          
    \end{aligned}
\end{equation}
and $S_q^{ab}(x)$ are the quark ($q=s,c$) propagators. For the light quark ($q=s$), propagators are expressed in terms of~\cite{Huang:2010dc,Agaev:2016mjb}
\begin{eqnarray} \nonumber
        S_{q}^{ab}(x)&=&\frac{i\delta_{ab}\slashed{x}}{2\pi^2x^4}-\frac{\delta_{ab}m_q}{4\pi^2x^2}-\frac{{\langle \bar{q}q\rangle }}{12}\\\nonumber
        &&-\frac{i}{32\pi^2}\frac{\lambda^e}{2}g_sG_{\mu\nu}^e\frac{1}{x^2}(\sigma^{\mu\nu}\slashed{x}+\slashed{x}\sigma^{\mu\nu})\\\nonumber
        &&+\frac{i\delta_{ab}\slashed{x}m_q{\langle \bar{q}q\rangle }}{48}
        -\frac{\delta_{ab}{\langle \bar{q}g_s\sigma Gq \rangle}x^2}{192} \\\nonumber
        &&+\frac{i\delta_{ab}x^2\slashed{x}m_q\langle\bar{q}g_s\sigma Gq\rangle}{1152}\\
        &&-\frac{i\delta_{ab}x^2\slashed{x}g_s^2\langle\bar{q}q\rangle^2}{7776}-\frac{\delta_{ab}x^4\langle\bar{q}q\rangle\langle g_s^2GG\rangle}{27648}, \quad\quad
\end{eqnarray}
here $e$ is the color index,
and the heavy quark ($q=c$) propagator is given by \cite{Reinders:1984sr,Yang:2020wkh}
\begin{eqnarray}\label{Hpropagator} \nonumber
S_{q}^{a b}(x)&=& i \int \frac{d^{4} k}{(2 \pi)^{4}} e^{-i k x}\left\{ \frac{\delta_{a b}\left(k+m_{q}\right)}{k^{2}-m_{q}^{2}} \right. \\ \nonumber
&& \left. -\frac{g_{s} G^{a b}_{\mu \nu}}{4} \frac{\sigma_{\mu \nu}\left(k+m_{q}\right)+\left(k+m_{q}\right) \sigma_{\mu \nu}}{\left(k^{2}-m_{q}^{2}\right)^{2}} \right. \\
  && \left. +\frac{g_{s}^{2} G^{2}}{12} \delta_{a b} m_{q} \frac{k^{2}+m_{q} k}{\left(k^{2}-m_{q}^{2}\right)^{4}} +\cdots\right\}.
\end{eqnarray}
Here we indicate that
\begin{equation}
\begin{aligned}
G^{a b}_{\mu \nu} \equiv G^{f}_{\mu \nu} t_{a b}^{f}, \quad G^{2}=G_{\mu \nu}^{f} G^{f \mu \nu} .
\end{aligned}
\end{equation}
where $f$ denote the color index. 

As we see, in the phenomenological side expressed in Eq.\eqref{Three-phenoDD},Eq.\eqref{Three-phenoDsDs}, there emerge several structures like $g_{\mu\nu}$, $p_\mu p_\nu$ and so on. 
We consider the $g_{\mu\nu}$ and $q_\mu$ structure in Eq.\eqref{Three-phenoDD}, Eq.\eqref{Three-phenoDsDs} respectively,  and make a Borel transform with respect to $-p^{2}=-q^2\to M^2$ on both the phenomenological and the OPE sides to get the coupling constant\cite{Dias:2013xfa}:  Hence, we derive:
\begin{eqnarray} \label{expresssion}
      &A(\frac{4 m_{D_s^*}^2-m_X^2}{M^2}e^{\frac{m_{D_s^*}^2}{M^2}}-4 e^{\frac{m_{D_s^*}^2}{M^2}}+4e^{\frac{m_X^2}{4 M^2}}) + B e^{-\hat{s}_0/M^2}\nonumber\\
      &=\Pi_1^{DsDs}(M^2)+\Pi_2^{DsDs}(M^2)
\end{eqnarray}
\begin{eqnarray} \label{expresssion1}
      &C(\frac{4 m_{D_s}^2-m_X^2}{M^2}e^{\frac{m_{D_s}^2}{M^2}}-4 e^{\frac{m_{D_s}^2}{M^2}}+4e^{\frac{m_X^2}{4 M^2}}) + D e^{-\hat{s}_0/M^2}\nonumber\\
      &=\Pi_1^{DD}(M^2)+\Pi_2^{DD}(M^2)
\end{eqnarray}
\begin{eqnarray} \label{expresssion2}
      &E(\frac{4 m_{D_s^*}^2-m_X^2}{M^2}e^{\frac{m_{D_s^*}^2}{M^2}}-4 e^{\frac{m_{D_s^*}^2}{M^2}}+4e^{\frac{m_X^2}{4 M^2}}) + F e^{-\hat{s}_0/M^2}\nonumber\\
      &=\Pi_1^{DDs}(M^2)+\Pi_2^{DDs}(M^2)
\end{eqnarray}
where $\hat{s}_0$ is the continuum threshold parameter for $X(4500)$,
\begin{eqnarray} \label{A}
      &&A=\frac{g_{XD_s^*D_s^*}f_{D_s^*}^2f_X}{(4m_{D_s^*}^2 - m_X^2)^2},C=\frac{g_{XD_sD_s}f_{D_s}^2f_Xm_{D_s}^4}{(m_c+m_s)^2(4m_{D_s}^2 - m_X^2)^2}\nonumber\\
      &&E=\frac{g_{XD_sD_s^*}f_{D_s^*}f_{D_s}f_Xm_{D_s}^2m_{D_s^*}}{(m_c+m_s)(4m_{D_s^*}^2 - m_X^2)^2}
\end{eqnarray}
and $B,D,F$ are the parameters introduced to take into account single pole contributions associated with pole-continuum
transitions in a three-point function sum rule \cite{Navarra:2006nd,Colangelo:1994es,Colangelo:1994es,Ioffe:1983ju}.
We show the details of $\Pi_1(M^2), \Pi_2(M^2)$ in Appendix \ref{appendix:C}.

To determine the coupling constant $g_{XD_s^*D_s^*}$ we can fit the
results with the analytical expression in the left-hand side with right-hand side of Eq.\eqref{expresssion} to find the value of 
A. Using the definition of A in Eq.\eqref{A}, we can obtain the values of coupling constants.

Finally, the decay width of $X(4500)$ $\to$ $A B$ can be calculated by equation \cite{Dias:2013xfa}
\begin{eqnarray}\label{width} \nonumber
        &&\Gamma(X(4500)\to A B)= \frac{(g_{XA B})^2}{24\pi m_{X}^2}\\
        &\times& \lambda(m_X,m_{A},m_{B})\left(3+\frac{\lambda(m_X,m_{A},m_{B})}{m^2_{A}}\right), \quad
\end{eqnarray}
where
\begin{eqnarray}
    \lambda(a,b,c)=\frac{\sqrt{a^4+b^4+c^4-2(a^2b^2+b^2c^2+c^2a^2)}}{2a}.
\end{eqnarray}

\subsection{The strong coupling \texorpdfstring{$g_{X J/\psi \phi}$}{Lg} in the three-point sum rules}

Next, we are going to predict the decay width of $X(4500)$ $\to$ $J/\psi\phi$.
We need to calculate the strong coupling $g_{X J/\psi \phi}$ in the first place.
The starting point is to write down the correlation function:
\begin{eqnarray}\label{three-point}
&&\Pi_{\mu \nu}^{\text{TP,Hidden}}\left(p^{\prime}, p, q\right) \\ \nonumber
&=&\int d^{4} x d^{4} y e^{i p \cdot x} e^{i q \cdot y}\left\langle 0\left|T\left[J^{J/\psi}_\mu(x) J^{\phi\dagger}_\nu(y) J^{X \dagger}(0)\right]\right| 0\right\rangle,
\end{eqnarray}
where the interpolating currents for $J/\psi$, $\phi$ are given by:
\begin{eqnarray}\label{currents}\nonumber
		&J_{\mu}^{J/\psi}(x)&=\bar{c}_m(x)\gamma_\mu c_m(x),\\
		&J_{\nu}^{\phi}(x)&=\bar{s}_n(x)\gamma_\nu s_n(x),\nonumber
\end{eqnarray}
the subscripts $m$, $n$ refer to the color indices.

Next, we insert intermediate states for $X(4500)$, $J/\psi$, $\phi$ into Eq.\eqref{three-point} and write down the correlation function as
\begin{eqnarray}\label{Three-pheno} \nonumber
&& \Pi_{\mu \nu}^{\text{TP,phen}}\left(p^{\prime}, p, q\right) \\\nonumber
&=& \frac{g_{X J/\psi \phi}^{TP} f_{J/\psi} f_{\phi} f_{X}}{\left(p^{\prime 2}-m_{X}^{2}+i \epsilon\right)\left(p^{2}-m_{J/\psi}^{2}+i \epsilon\right)\left(q^{2}-m_{\phi}^{2}+i \epsilon\right)} \times \\ \nonumber
&&\left(g_{\mu \nu}-\frac{p^\prime_{\mu} p_{\nu}+p^\prime_{\mu} p^\prime_{\nu}}{m_{X}^{2}}-\frac{p_{\mu} p_{\nu}}{m_{J/\psi}^{2}}+\frac{p^\prime \cdot p\left(p_{\mu} p_{\nu}+p^\prime_{\nu} p_{\mu}\right)}{m_{X}^{2} m_{J/\psi}^{2}}\right) \\
&&+\ldots.
\end{eqnarray}
In order to obtain Eq.\eqref{Three-pheno}, we have used the relationships:
\begin{eqnarray}\label{decay-cons}\nonumber
        &&\braket{0|J_{\nu}^{\phi}| \phi(q) }=m_{\phi}f_{\phi}\varepsilon_{\nu}^\prime,\\ \nonumber
        &&\braket{0|J_{\mu}^{J/\psi}|J/\psi(p)}=m_{J/\psi}f_{J/\psi}\varepsilon_\mu,
\end{eqnarray}
in which $\epsilon^\prime_\nu$, $\epsilon_\mu$ are the polarization vectors of $\phi$ and $J/\psi$ respectively.
$f_{\phi (J/\psi)}$ is the decay constant of $\phi (J/\psi)$.
Besides, the coupling constant $g_{XJ/\psi\phi}^{TP}$ came from the matrix element
$\langle\phi(q)J/\psi(p)|X(p^\prime)\rangle$:
\begin{equation}\label{coupling}
    \begin{aligned}
        \langle\phi(q)J/\psi(p)|&X(p^\prime)\rangle\\
        =&g_{XJ/\psi \phi}^{TP}
        [(q\cdot \varepsilon^*)(p\cdot\varepsilon^\prime)-(q\cdot p)(\varepsilon^*\cdot \varepsilon^\prime)].
    \end{aligned}
\end{equation}

After applying the Wick Theorem on Eq.\eqref{three-point}, the OPE side of the sum rules is:
\begin{eqnarray} \nonumber
		&& \Pi_{\mu \nu}^{\text{TP,Hidden,OPE}}\left(p^{\prime}, p, q\right) =\int d^{4} x d^{4} y e^{i p \cdot x} e^{i q \cdot y} \\\nonumber
		&&\{\text{Tr}[S^{nl}_s(-y)\gamma_\nu S^{nl}_s(y-z)\overleftarrow{D}_{\mu 3}\overleftarrow{D}_{\mu 4} \\ \nonumber
		&&~~~~~~\gamma_{\nu1} \tilde{S}^{km}_c(x)\gamma_\mu\tilde{S}^{km}_c(-x)\gamma_{\mu2}]\\\nonumber
		&&-\text{Tr}[S^{nk}_s(-y)\gamma_\nu S^{nl}_s(y-z)\overleftarrow{D}_{\mu 3}\overleftarrow{D}_{\mu 4} \\
		&&~~~~~~\gamma_{\nu1}\tilde{S}^{km}_c(x)\gamma_\mu\tilde{S}^{lm}_c(-x)\gamma_{\mu2}] \} ,
\end{eqnarray}
We consider the $p^\prime_\mu p^\prime_\nu$ structure in this works.
Following Refs.\cite{Dias:2013xfa,Navarra:1998vi,Bracco:1999xe}, we will neglect $m_\phi^2$ in the denominators. 
Therefore, only those terms proportional to $\frac{1}{q^2}$ will contribute
to the OPE side ~\cite{Choe:1998zi,Choe:1995yb,Reinders:1984sr}. Hence, we derive the spectral density as:
\begin{eqnarray} 
		\rho^{TP,Hid}(\hat{s},Q^2)&=& \frac{9 \braket{\frac{\alpha_s}{\pi}GG} m_c^2 \sqrt{\hat{s} \left(\hat{s}-4 m_c^2\right)}}{16 \pi^2 Q^2 \hat{s} \left(\hat{s}-4 m_c^2\right)^2}(1-\frac{2m_c^2}{\hat{s}}) ~~~~~~
\end{eqnarray}
where $Q^2=-q^2$ and
\begin{eqnarray} 
        \rho^{\text{TP,Hid}}(\hat{s},Q^2)=\frac{\text{Im}\Pi^{\text{TP,OPE}}(p^\prime,p,q)}{\pi}.
\end{eqnarray}
Finally, we make a Borel transform with respect to $P^{\prime2}=-p^{\prime2}=p^2\to M^2$ on both the phenomenological
and the OPE sides to get the coupling constant\cite{Dias:2013xfa}: 
\begin{eqnarray} \label{Three-Point} \nonumber
&&\hat{g}_{XJ/\psi\phi}^{TP}\left(\hat{s}_{0}, M^{2}, Q^{2}\right)= \\ \nonumber
&& \frac{1}{\lambda_{\phi}\lambda_{J/\psi} \lambda_{X}} \frac{m_{X}^{2}-m_{J/\psi}^{2}}{e^{-m_{J/\psi}^{2} / M^{2}}-e^{-m_{X}^{2} / M^{2}}} \\
&& \times\left(Q^{2}+m_{J/\psi}^{2}\right)\int_{4 (m_{c}^{2}+m_s^2)}^{\hat{s}_{0}} \rho^{\text{TP}}(\hat{s},Q^2) e^{-\hat{s} / M^{2}} d \hat{s} .~~~~
\end{eqnarray}
Here $\hat{s}_0$ is the continuum threshold
parameter for $X(4500)$ and $M^2$ is the Borel parameter for $p^2$.
Finally, the decay width of $X(4500)$ $\to$ $J/\psi \phi$ can be calculated by equation\eqref{width}.

\subsection{The strong coupling \texorpdfstring{$g_{X J/\psi \phi}$}{Lg} in LCSR}

Under the Techne of three-point sum rules, $g_{X J/\psi\phi}$, as a form factor, 
would be portrayed as some vacuum expectation values of operator involving quark fields.
Besides three-point sum rules, we can also describe $g_{X J/\psi\phi}$ in the framework of LCSR where
the coupling will quantify with the quark distribution in longitudinal momenta inside the hadron.
So, as a double check in this section, we employ the light-cone sum rules method to calculate $g_{X J/\psi\phi}$.
Imprimis, we consider the correlation function
\begin{equation}\label{3}
	\begin{aligned}
	\Pi_{\mu}^{\text{LC}}(p^\prime,p,q)=i\int \mbox{d}^4xe^{ipx}\braket{\phi(q)|T\{J_{\mu}^{J/\psi}(x)J^{X\dagger}(0)\}|0}.
	\end{aligned}
\end{equation}

Similarly, by putting the intermediate into itself, the phenomenological expressions of the correlation function can be written as follows:
\begin{equation}\label{pheno}
	\begin{aligned}
		&\Pi^{\text{LC,phen}}_{\mu}(p^\prime,p,q)=\\
		&\frac{\braket{0|J_{\mu}^{J/\psi}|J/\psi(p)}
		       \braket{\phi(q)J/\psi(p)|X(p^\prime)}
			 \braket{X(p^\prime)|J^{{X}\dagger}|0}}
			 {(p^{\prime2}-m_{X}^2)(p^2-m_{J/\psi}^2)}\\
			 &+\cdots.
	\end{aligned}
\end{equation}
Here the dots denote contributions of the higher resonances states.
Now, by parameterizing the above equation with Eq.\eqref{decay-cons}, and introducing the hadronic matrix element
\begin{equation}\label{coupling1}
	\begin{aligned}
		\langle\phi(q)J/\psi(p)|&X(p^\prime)\rangle\\
		=&g_{XJ/\psi \phi}^{LC}
		[(q\cdot \varepsilon^*)(p\cdot\varepsilon^\prime)-(q\cdot p)(\varepsilon^*\cdot \varepsilon^\prime)],
	\end{aligned}
\end{equation}
we express the phenomenological side of the correlation function \eqref{pheno} as
\begin{equation}
	\begin{aligned}\label{1}
		\Pi^{\text{LC,phen}}_{\mu}(p^\prime,p,q)&=\frac{m_{J/\psi}m_{X}f_{J/\psi}f_{X}g_{XJ/\psi \phi}^{LC}}{(p^{\prime2}-m_{X}^2)(p^2-m_{J/\psi}^2)}\\
		&\times [ (p\cdot q)\epsilon_{\mu}^\prime-p\cdot\epsilon^\prime q_\mu ] +\cdots\\
		&=\Pi^{\text{Phys}}(p^\prime,q) [(p\cdot q)\epsilon_{\mu}^\prime-p\cdot\epsilon^\prime q_\mu ],
	\end{aligned}
\end{equation}
where the $g_{XJ/\psi \phi}^{LC}$ is the coupling constant of $X(4500)\to J/\psi\phi$ in LCSR.

Next, we select the structure that is proportional to $ (p\cdot q)\epsilon_{\mu}^\prime-p\cdot\epsilon^\prime q_\mu $ and write the relevant structure down as
\begin{equation}
	\begin{aligned}
		\Pi^{\text{LC,phen}}(p^\prime,p,q)&=\frac{m_{J/\psi}m_{X}f_{J/\psi}f_{X}g_{XJ/\psi \phi}^{LC}}{(p^{\prime2}-m_{X}^2)(p^2-m_{J/\psi}^2)}\\
		&+\int_{s_1^{0\prime}}^\infty\int_{s_2^{0\prime}}^\infty\frac{\mbox{d}s_1\mbox{d}s_2\rho^{\text{LC,phen}}(s_1,s_2)}{(s_1-p^2)(s_2-p^{\prime2})}+\cdots.
	\end{aligned}
\end{equation}
Where $\rho^{\text{LC,phen}}(s_1,s_2)$ represents contributions of the higher resonances and the continuum states.
Now, what we have to do is to perform the Borel transformations to the correlation function, and express the result as
\begin{equation}\label{ha}
	\begin{aligned}
		&\mathcal{B}_{p^2}(M_1^2)\mathcal{B}_{p^{\prime 2}}(M_2^2)\Pi^{\text{LC,phen}}(p^\prime,p,q)=\\
		&m_{J/\psi}m_{X}f_{J/\psi}f_{X}g_{XJ/\psi \phi}^{LC}\exp[-\frac{m_{J/\psi}^2}{M_1^2}-\frac{(m_{X})^2}{M_2^2}]\\
		&+\int_{s_1^{0\prime}}^\infty\int_{s_2^{0\prime}}^\infty\mbox{d}s_1\mbox{d}s_2 \exp[-\frac{s_1}{M_1^2}-\frac{s_2}{M_2^2}] \rho^{\text{LC,phen}}(s_1,s_2).
	\end{aligned}
\end{equation}

Phenomenological sides of LCSR gives us connection between hadron property and the function \eqref{3}. Then we focus on the OPE side that connect the function \eqref{3} to the quark longitudinal distribution. First, we express the correlation function as a double dispersion integral:
\begin{equation}
	\begin{aligned}
	\Pi^{\text{LC,OPE}}(p^\prime,p,q)
	=\int_{s_1^{\prime}}^\infty\int_{s_2^{\prime}}^\infty \frac{ds_1ds_2\ \rho^{\text{LC,OPE}}(s_1,s_2)}{(s_1-p^2)(s_2-p^{\prime2})}
	+\cdots,
	\end{aligned}
\end{equation}
where the term $\rho^{\text{OPE}}(s_1,s_2)$ is generally understood as
\begin{equation}
	\begin{aligned}
	\rho^{\text{LC,OPE}}(s_1,s_2)=\frac{\text{Im}^{\text{LC,OPE}}\Pi(s_1,s_2)}{\pi^2}.
	\end{aligned}
\end{equation}
In generally, $\text{Im}\Pi(s_1,s_2)$ include the quark distribution function.

Our purpose is to obtain the connection between the Phenomenological sides and the OPE sides of LCSR. The main idea is to use the quark-hadron duality\cite{Braun:2012kp}, which allows one to express the coupling as
\begin{equation}\label{h}
	\begin{aligned}
	&\tilde{g}_{XJ/\psi \phi}^{LC}=\frac{1}{m_{J/\psi}m_{X}f_{J/\psi}f_{X}}\exp[\frac{m_{J/\psi}^2}{M_1^2}
	+\frac{(m_{X})^2}{M_2^2}]\\
	&\times\int^{s_1^0}_{s_1^\prime}\int^{s_2^0}_{s_2^\prime}\mbox{d}s_1\mbox{d}s_2
	\exp[-\frac{s_1}{M_1^2}-\frac{s_2}{M_2^2}] \rho^{\text{LC,OPE}}(s_1,s_2).
	\end{aligned}
\end{equation}

However, there is a difference between our situation and the standard one shown above.
As we can see from Eq.\eqref{3}, the interpolating current of $X(4500)$ belongs to the space-time point $0$,
and the interpolating current of $J/\psi$ belongs to the space-time point $x$.
Therefore the structure not $\braket{\phi(q)|[\bar{s}(x)s(0)]|0}$ but $\braket{\phi(q)|[\bar{s}(0)s(0)]|0}$
remains after contract the $\bar{c}$ and $c$ quark fields.
The structure $\braket{\phi(q)|[\bar{s}(0)s(0)]|0}$ cause the $\phi$ distribution reduces to normalization factor.
This situation only appear when $q\rightarrow 0$, and the correlation function now depends only on one variable $p^2$.
\begin{eqnarray}
		&\Pi^{\text{LC,phen}}(p)=\frac{m_{J/\psi}m_{X}f_{J/\psi}f_{X}}{(p^2-m^2)^2} g_{XJ/\psi \phi}^{LC}
			   +\cdots,
\end{eqnarray}
where $m^2=\frac{m_{J/\psi}^2+m_{X}^2}{2}$.
Notice $q\rightarrow 0$ simplifies the hadronic side of the sum rules, but leads to a more complicated expression on its hadronic representation.
Here, following Ref.\cite{Belyaev:1994zk}, by applying the Borel transformation on the variable $p^2$ to the correlation function, we rewrite the
phenomenological sides as
\begin{eqnarray} \nonumber
  \Pi^{\text{LC,phen}}(p)&=&\frac{1}{M^2}(m_{J/\psi}m_{X}f_{J/\psi}f_{X}g_{XJ/\psi \phi}^{LC} \\
   &&+A M^2)e^{\frac{-m^2}{M^2}}+C.
\end{eqnarray}
The coefficient  A involves all the unsuppressed contributions, 
while the coefficient C represents all the exponentially suppressed contributions.
To remove the unsuppressed parts, we conduct the following operator \cite{Ioffe:1983ju}
\begin{equation}
	\begin{aligned}
		(1-M^2\frac{d}{dM^2})M^2e^{m^2/M^2}
	\end{aligned}
\end{equation}
on both sides of the sum rules expressions
and come to the results of
\begin{equation}\label{couplingn}
	\begin{aligned}
		&\tilde{g}_{XJ/\psi \phi}^{LC}=\frac{1}{m_{J/\psi}m_{X}f_{J/\psi}f_{X}}(1-M^2\frac{d}{dM^2})M^2\\
		&\times\int^{\hat{s}^\prime}_{\hat{s}_0}\mbox{d}\hat{s}\exp[\frac{m_{J/\psi}^2}{2M^2}+\frac{(m_{X})^2}{2M^2}-\frac{\hat{s}}{M^2}] \rho^{\text{LC,OPE}}(\hat{s}).
	\end{aligned}
\end{equation}
The continuum state no longer depends on two variables $s_1$ and $s_2$, but rather on one variable labeled as $\hat{s}$ since we use the soft-meson approximation.

\subsection{The OPE side calculation in the LCSR}

Eq.\eqref{couplingn} connects the coupling $\tilde{g}_{XJ/\psi \phi}^{LC}$ to the OPE part of the correlation function \eqref{3}.
So we need to work out the OPE part of the correlation function in the LCSR.

Expressing the equation \eqref{3} with interpolating currents in Eq.\eqref{currents} and using the Wick Theorem, the OPE side of the correlation function is:
\begin{equation}\label{31}
	\begin{aligned}
	\Pi_{\mu}^{\text{LC,OPE}}(p^\prime,p,q)
	&=i\int d^4xe^{ipx}\braket{\phi(q)|T\{J_{\mu}^{J/\psi}(x)J^{X\dagger}(0)\}|0}\\
	&=i\int d^4xe^{ipx}\{\braket{\phi(q)|[(\bar{s}^l(0)\overleftarrow{D}_{\mu 4}\overleftarrow{D}_{\mu 3})_\alpha s^l_\beta(0)]|0}\\
	&\times[\gamma_{\mu 1}\tilde{S}^{km}_c(x)\gamma_\nu\tilde{S}^{km}_c(-x)\gamma_{\mu 2}]_{\alpha\beta}\\
	&-\braket{\phi(q)|[(\bar{s}^l(0)\overleftarrow{D}_{\mu 4}\overleftarrow{D}_{\mu 3})_\alpha s^k_\beta(0)]|0}\\
	&\times\gamma_{\mu 1}\tilde{S}^{km}_c(x)\gamma_\nu\tilde{S}^{lm}_c(-x)\gamma_{\mu 2}]_{\alpha\beta}\}.
	\end{aligned}
\end{equation}
It is necessary to rule out the index in the next step. So we introduce the expansion
\begin{equation}\label{summation}
\begin{aligned}
\bar{s}^d_\alpha(0) s^{d^\prime}_\beta(0)=\frac{1}{12}\delta_{dd^\prime}\Gamma^a_{\alpha\beta}\bar{s}(0)\Gamma^a s(0),
\end{aligned}
\end{equation}
where $\Gamma^a=1,\gamma_5,\gamma_\mu,i\gamma_5\gamma_\mu,\frac{\sigma_{\mu\nu}}{\sqrt{2}}$.

After performing the replacement of Eq.\eqref{summation} in Eq.\eqref{31}, the correlation function can be rewritten as follows:
\begin{equation}\label{opecorrelation1}
	\begin{aligned}
	\Pi_{\mu}^{\text{LC,OPE}}(p^\prime,p,q)
	&=i\int d^4xe^{ipx}\braket{\phi(q)|T\{J_{\mu}^{J/\psi}(x)J^{X\dagger}(0)\}|0}\\
	&=i\int d^4xe^{ipx}\{\braket{\phi(q)|[(\bar{s}^k(0)\overleftarrow{D}_{\mu 4}\overleftarrow{D}_{\mu 3})\Gamma^e s^k(0)]|0}\\
	&\times\text{Tr}[\gamma_{\mu 1}\tilde{S}^{ji}_c(x)\gamma_\nu\tilde{S}^{ji}_c(-x)\gamma_{\mu 2}]\\
	&+\braket{\phi(q)|[(\bar{s}^k(0)\overleftarrow{D}_{\mu 4}\overleftarrow{D}_{\mu 3})\Gamma^e s^j(0)]|0}\\
	&\times\text{Tr}[\gamma_{\mu 1}\tilde{S}^{ji}_c(x)\gamma_\nu\tilde{S}^{ki}_c(-x)\gamma_{\mu 2}]\}.
	\end{aligned}
\end{equation}

The matrix element $\braket{\phi(q)|[(\bar{s}^k(0)\overleftarrow{D}_{\mu 4}\overleftarrow{D}_{\mu 3})\Gamma^e s^k(0)]|0}$ relate to the
so-called $\phi$ distribution amplitudes(DAs), which is given in \ref{appendix:A}.
After replacing the propagator by Eq.\eqref{Hpropagator} and employing the DAs of $\phi$,
we find that there are four-dimensional integrals in the momentum spaces that appear in Eq.\eqref{opecorrelation1}. For instance
\begin{equation}\label{four integral}
	\begin{aligned}
		\int\frac{d^4k_1}{(2\pi)^4}\int\frac{d^4k_2}{(2\pi)^4}\frac{e^{-i(k_1-k_2)x} k_1\cdot k_2}{(k_1^2-m^2_c)(k_2^2-m^2_c)}[(p\cdot q)\epsilon_{\mu}^\prime-p\cdot\epsilon^\prime q_\mu].
	\end{aligned}
\end{equation}
We can calculate those integrals in D dimension, and then dimensionally regularized at D = 4 \cite{Matheus:2006xi}.
In the results, the infinity only comes from the real part, and because the spectral density just relates to the imaginary part,
so spectral density will be a finite result.
By selecting the structure proportional to $(p\cdot q)\epsilon_{\mu}^\prime-p\cdot\epsilon^\prime q_\mu$, we derive the associated spectral density.
\begin{equation}\label{OPER}
	\begin{aligned}
		\rho^{\text{LC,OPE}}(\hat{s})=\int_0^1d uu^2\phi_2^{\bot}(u)\frac{\sqrt{2}m_c m_{\phi}^2f_{\phi}^{\bot}\sqrt{\hat{s}(\hat{s}-4m_c^2)}}{4\pi^2\hat{s}},
	\end{aligned}
\end{equation}
where $m_c$ denote the charm quark mass. The mass and decay constant of $\phi$ are represented by $m_\phi$ and $f_\phi^\bot$, respectively.  
$\phi_2^{\bot}(u)$ is the light cone distribution amplitudes of $\phi$. 

\begin{figure}[H]
\centering
  \includegraphics[width=5cm]{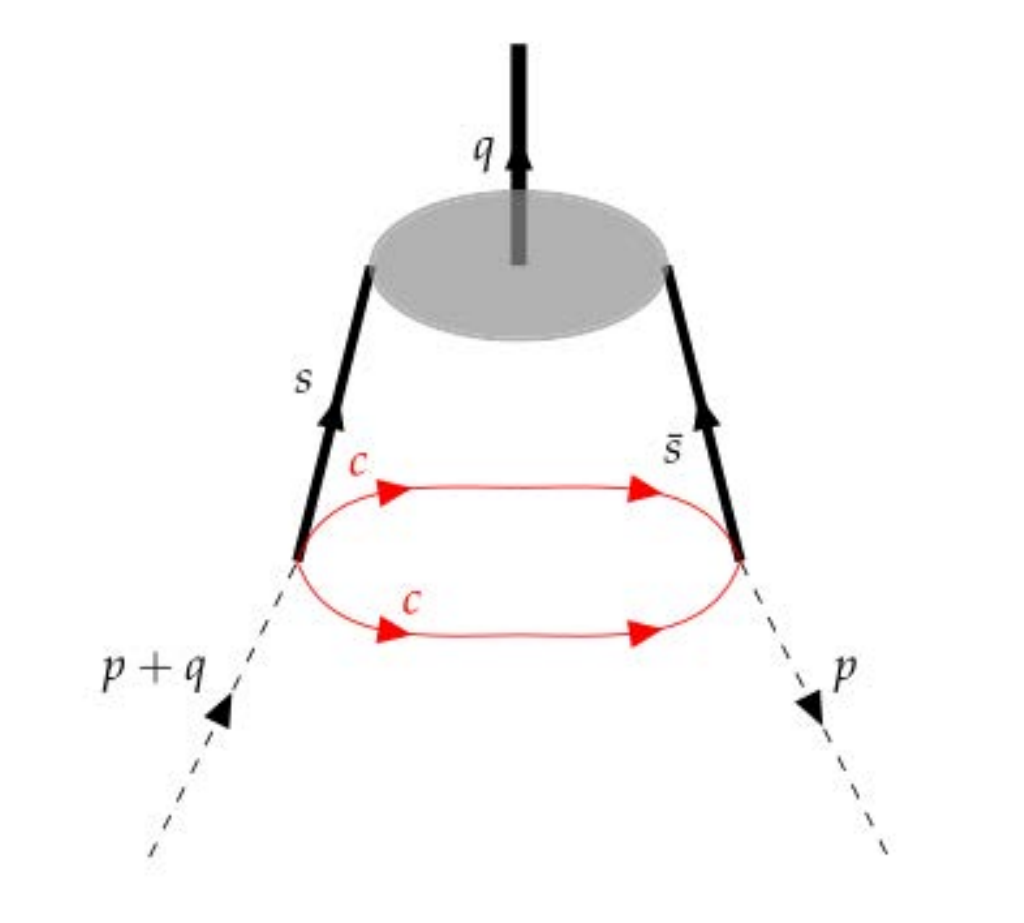}
  \caption{The leading order diagram contribute to $\Pi_{\mu}(p^\prime,p,q)$.}
\label{Fig:fig1a}
\end{figure}
\begin{figure}[htpb]
\centering
  \includegraphics[width=5cm]{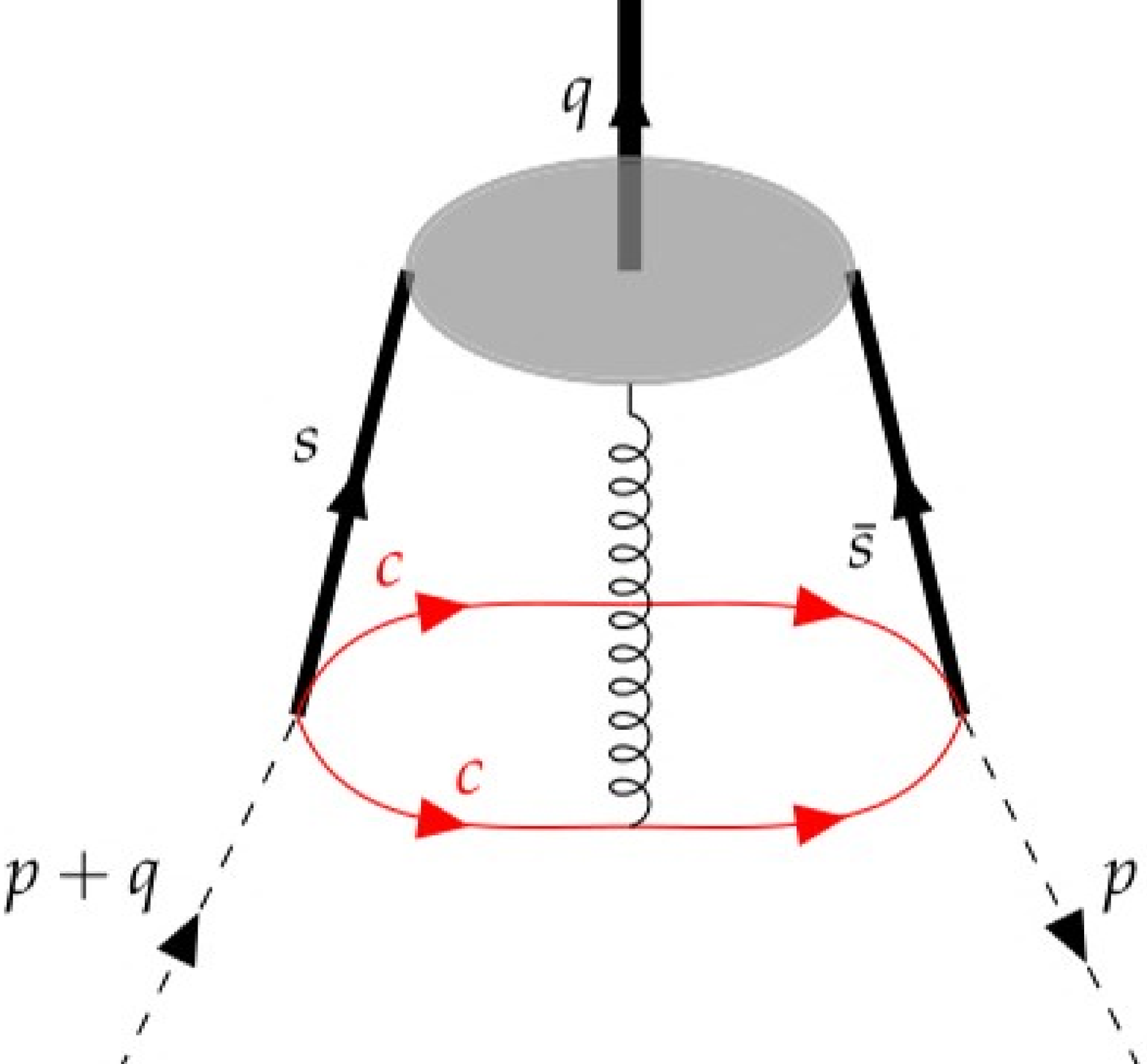}
  \caption{The one-gluon exchange contribution to $\Pi_{\mu}(p^\prime,p,q)$.}
\label{Fig:fig1b}
\end{figure}
FIG.\ref{Fig:fig1a} represents the Feynman diagram of $\Pi_{\mu}(p^\prime,p,q)$,
represents the leading order contribution, which is dominant over the one-gluon exchange contribution in FIG.\ref{Fig:fig1b}, for instance.
Therefore in Eq.\eqref{OPER}, we only keep the leading order contribution.

Now, it is straightforward that one can evaluate the strong coupling by Using Eq.\eqref{couplingn},
and hence the decay width of $X(4500)$ $\to J/\psi \phi$ can be calculated through Eq.\cite{Dias:2013xfa}
 \begin{equation}
 	\begin{aligned}
 		&\Gamma^{\text{LC}}(X(4500)\to J/\psi \phi)=\frac{(\tilde{g}_{XJ/\psi \phi}^{LC})^2}{24\pi m_{X}^2}\\
 		&\times\lambda(m_X,m_{J/\psi},m_{\phi})\left(3+\frac{\lambda(m_X,m_{J/\psi},m_{\phi})}{m^2_{J/\psi}}\right).
 	\end{aligned}
 \end{equation}

\section{Numerical calculation}
\label{III}

\subsection{Input parameters}

After theoretical preparation, we commence the numerical calculations 
for the mass and the decay constant of $X(4500)$, and for the decay width of $X(4500) \to J/\psi \phi$ as well.

First, the values of the non-perturbative vacuum condensates are presented as \cite{Wang:2016gxp}
\begin{equation}
	\begin{aligned}
		&\langle{\bar{q}q}\rangle=-(0.24\pm 0.01)^3\ \text{GeV}^3,\\
		&\braket{\bar{s}s}=(0.8\pm0.1)\times \braket{\bar{q}q},\\
		&\braket{g_s\bar{s}\sigma Gs}=m_0^2\times\braket{\bar{s}s},\\
		&m_0^2=0.8\ \text{GeV}^2,\\
		&\braket{\frac{\alpha_s}{\pi} GG}=(0.012)\ \text{GeV}^4\added{.}\\
	\end{aligned}
\end{equation}

Second, we adopt the decay constants of $\phi$ and $J/\psi$ as $f_{\phi}^{\|}=0.215$ GeV \cite{Ball:2007zt} and $f_{J/\psi}=0.405$ GeV \cite{Dias:2013xfa} respectively. Meanwhile,
from Particla Data Group (PDG) \cite{ParticleDataGroup:2020ssz}, the current-quark-mass for the s-quark and charm-quark are taken as $m_s=93^{+11}_{-5}$ MeV and $m_c=(1.275\pm 0.025)$ GeV respectively, and, likewise, we accept the $J/\psi$-meson mass $m_{J/\psi}=(3096.900\pm 0.006)$ MeV and the $\phi$ mass $m_{\phi}=(990\pm 20)$ MeV.
The parameters $\zeta_4^{\bot}$ and $\widetilde{\zeta}_4^{\bot}$ are taken as $\zeta_4^{\bot}=-0.01$ and $\widetilde{\zeta}_4^{\bot}=-0.03$ \cite{Ball:2007zt}. 
Besides, for the Gegenbauer moments, we take $a_1^{\|}=a_1^{\bot}=0$, $a_2^{\|}=0.18$ and $a_2^{\bot}=0.14$ \cite{Ball:2007zt}.

\subsection{The mass and the decay constant }

Since we consider $X$(4500) as a D-wave tetraquark state, the primary thing we need to know is whether our conjecture is tenable or not.
So, before we evaluate the coupling, we should first evaluate the mass of $X(4500)$.
In addition, we also need to figure out the decay constant of $X(4500)$ because it is an indispensable parameter for coupling.

There are two parameters that the sum rules predictions will depend on: the Borel mass $M^2$ and the continuum threshold $s_0$.

As a first step, $s_0$ can be determined by two principles:
\begin{enumerate}
	\item 
	The mass value of the considered hadron should rely on $s_0$ as weak as possible.
	\item 
	The $s_0$ should be related to the first excited states of hadron.
\end{enumerate}
Because of the absence of experimental data of $X(4500)$, we cannot use the second principle directly.
Although, one may naturaly chose $s_0=(m_X+0.5)^2$ $\rm GeV^2$,
because the mass gap between the ground state and the first excited state is regularly around $0.5$ GeV in charmonia
and bottomonia~\cite{Olpak:2016wkf,ParticleDataGroup:2012pjm}, 
it may, however, not be true for $X(4500)$. 
So, the first principle become our primary choice.

We therefore define the function of
\begin{eqnarray}
  \Delta(M^2,s_0)=\left (\frac{\partial m_X}{\partial (M^2)}\right )^2+\left (\frac{\partial m_X}{\partial (s_0)} \right )^2
\end{eqnarray}
to describe the variation degree of the mass. 
In addition, we also impose the following constrain on $s_0$:
\begin{equation}
	\begin{aligned}
		s_0<(m_X+1)^2\ \text{GeV}^2 ,
	\end{aligned}
\end{equation}
since the energy gap between the ground state and the first excited state is usually smaller than 1 GeV~\cite{Wu:2021tzo}.
Through the numerical calculation of $\Delta(M^2,s_0)$ with a varying range of $M^2$ and $ s_0$, 
we found that, in a large range of $M^2$, the value of $\Delta(M^2, s_0)$ is close enough to the minimal value when $s_0$ is around $(m_X+0.398)^2\ \text{GeV}^2 $.
Therefore, for $X(4500)$, we employ
\begin{eqnarray} \label{Eq:s0}
  (4.90-0.10)^2\ \text{GeV}^2\leq s_0\leq (4.90+0.10)^2\ \text{GeV}^2.
\end{eqnarray}

Secondly, to determine the Borel mass $M^2$, we implement two criteria:
\begin{enumerate}
\item
The contribution proportional to $\braket{\bar{q}g_sGq}$ and higher dimension condensates should be less than $10\%$ of the gross contribution:
\begin{equation}
	\begin{aligned}
	\mathrm{CVG} \equiv\left|\frac{\Tilde{\Pi}^{\braket{\bar{q}g_sGq}+\cdots}\left( M^{2},\infty\right)}{\Tilde{\Pi}^{OPE}\left(M^{2},\infty \right)}\right| \leq 10 \% ,
	\end{aligned}
\end{equation}
where the dots means the terms with dimensions higher than $\braket{\bar{q}g_sGq}$.
\item The pole contribution (PC) should exceed $70\%$:
\begin{equation}
	\begin{aligned}
		\text{PC}=\frac{\Pi^{SVZ,OPE}(M^2,s_0)}{\Pi^{SVZ,OPE}(M^2,\infty)}\ge  70\% .
	\end{aligned}
\end{equation}
\end{enumerate}
Those two criteria settle the minimal and maximal values of $M^2$ respectively.

\begin{figure}[htbp]
\centering
  \includegraphics[width=6cm]{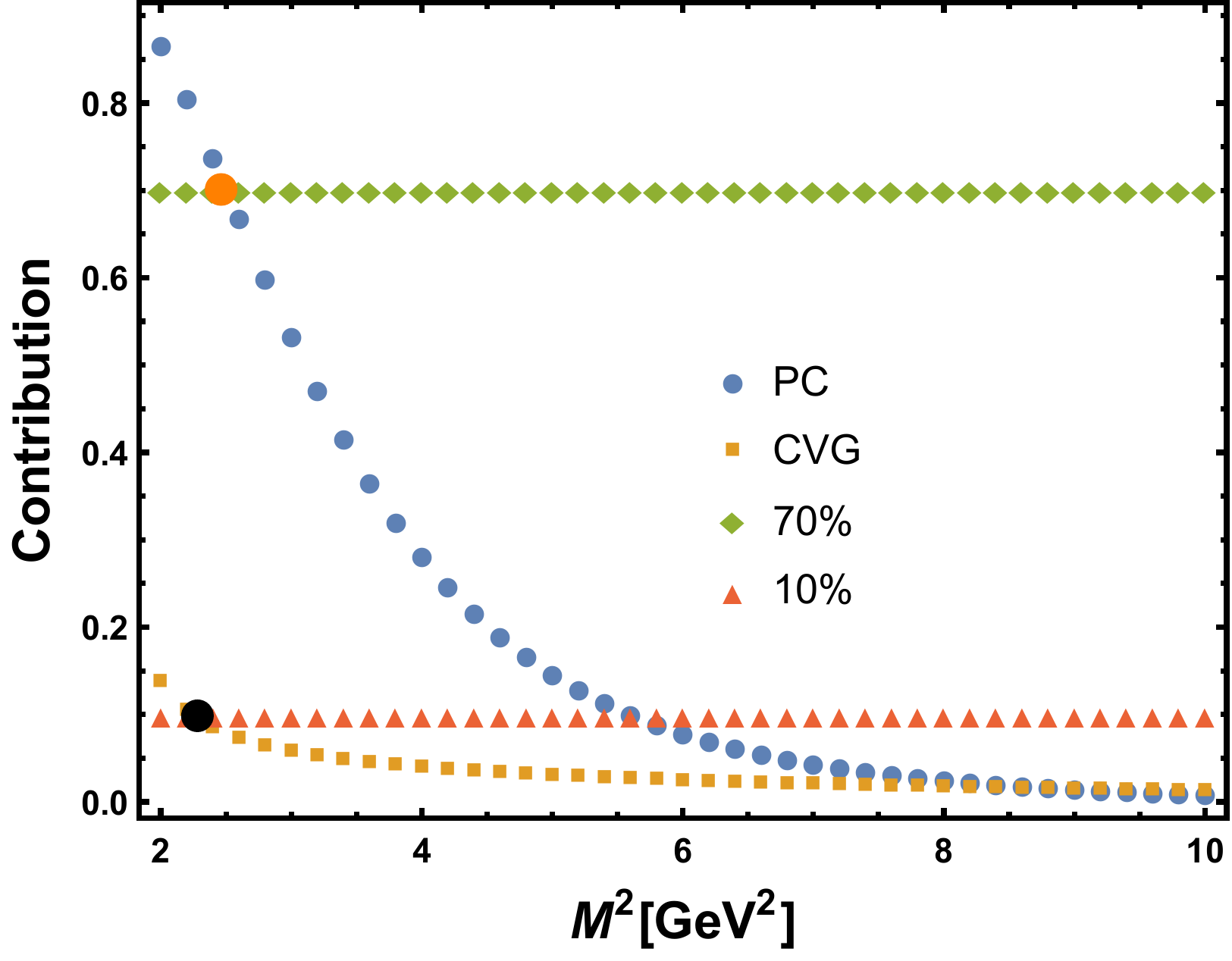}
  \caption{
  Convergence (CVG) and pole contribution (PC) for $X(4500)$. }
\label{Fig:fig2}
\end{figure}

The above criteria are shown in FIG.\ref{Fig:fig2}. 
CVG and PC are represented in the yellow and blue curve respectively,
they are both declining with the increase of $M^2$.
The black dot indicates that the CVG intersects horizontally with the $10\%$ line, from which we can choose the minimum value of the Borel mass. 
Similarly, the orange dot indicates that the CVG reaches $70\%$, and we can determine the maximum value of the Borel mass.
Therefore, the working region of the Borel mass turns to be
\begin{equation}
	\begin{aligned}
		2.28\ \text{GeV}^2\leq M^2\leq 2.46\ \text{GeV}^2.
	\end{aligned}
\end{equation}

\begin{figure}[htbp]
\centering
  \includegraphics[width=6cm]{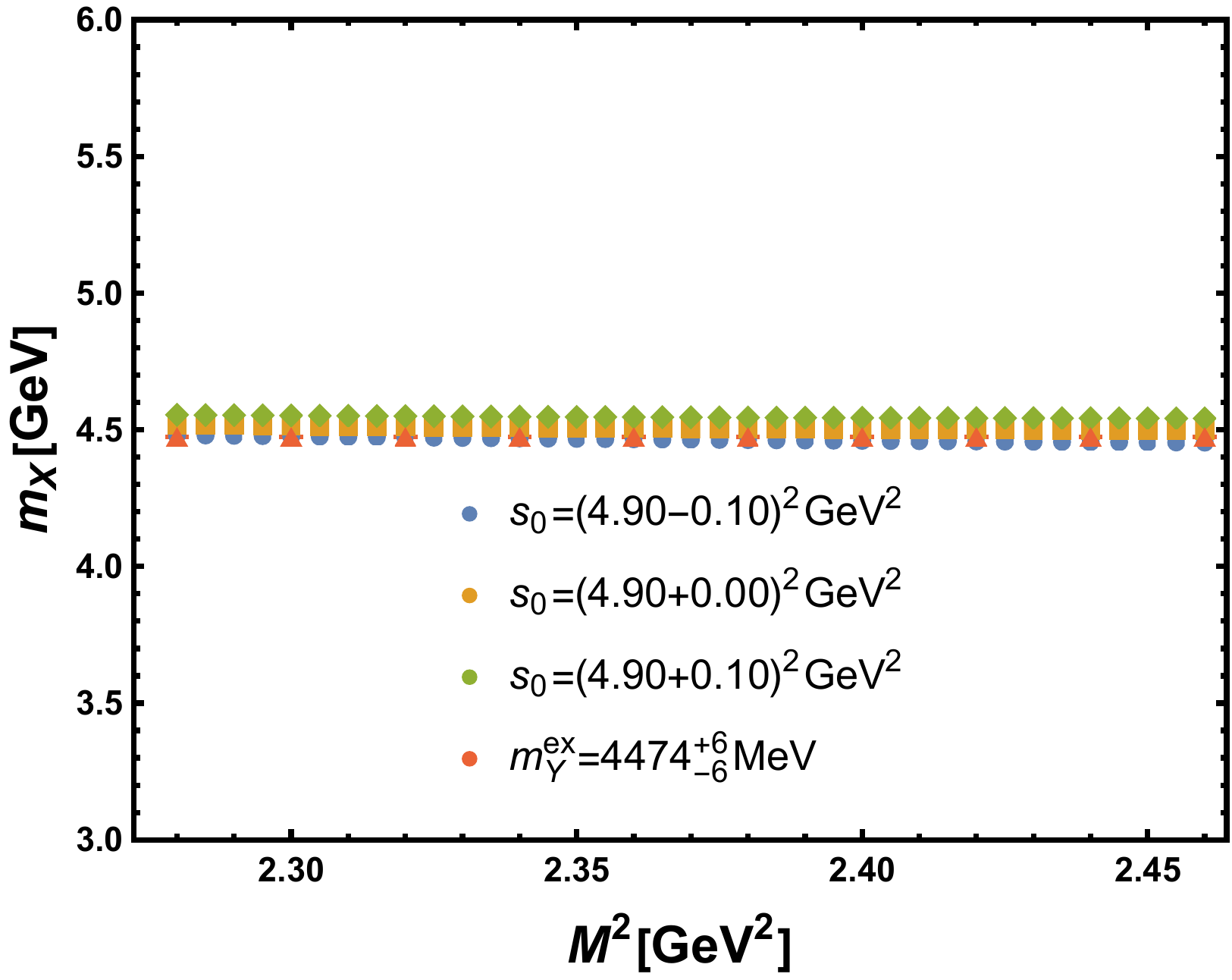}
  \includegraphics[width=6cm]{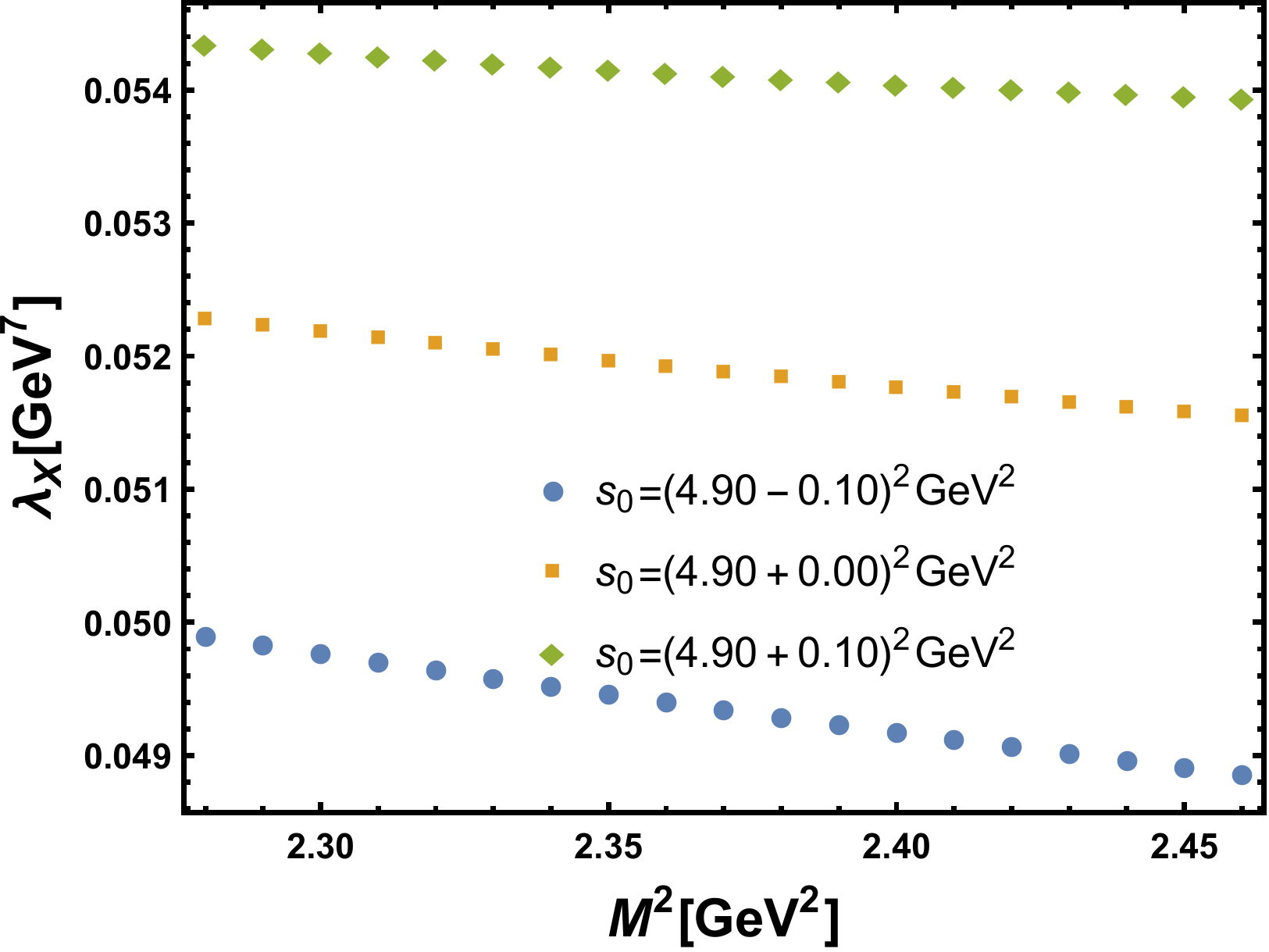}
  \caption{
 The mass [first] and the decay constant [second] of $X(4500)$ as a function of the Borel Mass $M^2$ at different fixed values of the continuum threshold $s_0$.}
\label{Fig:fig3}
\end{figure}

The calculation results of the mass and the decay constant have been depicted in Fig.\ref{Fig:fig3}.
We illustrate the results at fixed values of $s_0$$\in$$\{(4.90-0.10)^2\ \text{GeV}^2 $, $(4.90+0.0)^2\ \text{GeV}^2 $, $(4.90+0.1)^2\ \text{GeV}^2 \}$
and show them with the blue, orange and green curves respectively.
Base on the LHCb measurements~\cite{LHCb:2021uow}, $X(4500)$ has mass of $4474\pm 3\pm 3$ MeV.
As shown in the figure, the blue, orange and green curves overlap with the experimental value in the working region of the Borel mass.
At the central point of $M^2=2.37$ $\rm GeV^2$ and $s_0=(4.9+0.0)^2$ $\rm GeV^2$, the mass of $X(4500)$ can be extracted to be
\begin{equation}
	\begin{aligned}
		\quad  m_X=4.51^{+0.05}_{-0.04}\ \text{GeV}.
	\end{aligned}
\end{equation}
The uncertainty comes from the various condensates and the strange and charm quark masses. Our calculated results is consistent with the measurements of $X(4500)$.
So it's tenable that $X(4500)$ might be a D-wave $cs\bar{c}\bar{s}$ tetraquark.

Meanwhile, at the same benchmark point, our prediction of the decay constant is
\begin{equation}
	\begin{aligned}
	\lambda_X\equiv  m_X f_X=0.0521^{+0.0034}_{-0.0023}\ \text{GeV}^7.
	\end{aligned}
\end{equation}

Our next step is to calculate the decay width of $X$(4500)$\to J/\psi \phi$ using the mass and decay constant given above.

\subsection{The coupling constant and the decay width}

To determine the coupling constant $g_{XD_s^*D_s^*}$ we can fit the
 results with the analytical expression in the left-hand side with right-hand side of Eq[.\eqref{expresssion}] and find $A=-0.0063\  \text{GeV}^5, B=-5.3558\ \text{GeV}^5,\ C=-0.3850\text{GeV}^5, D=-3.6439\ \text{GeV}^5$ $E=0.0044\ \text{GeV}^5$, $F=0.4475\ \text{GeV}^5$. Using the definition in Eq.\eqref{A}, we can obtain the values of coupling constants
\begin{eqnarray}
g_{XD_sD_s}&=& -1.34884_{-0.083}^{+0.089}\ \text{GeV},\nonumber\\
 g_{XD_s^*D_s^*}&=& -1.44925^{+0.14}_{-0.15}\ \text{GeV},\\\nonumber
 g_{XD_sD_s^*}&=& 1.00197_{-0.165}^{+0.183}\ \text{GeV},\\
\end{eqnarray}
Therefor we obtain from Eq.\eqref{width}:
\begin{eqnarray}
 \Gamma^{\text{LC}}(X(4500) \rightarrow D_s D_s)= 4.28^{+0.54}_{-0.51} \ \text{MeV}\nonumber\\
    \Gamma^{\text{LC}}(X(4500) \rightarrow D_s^* D_s^*)= 3.32^{+0.61}_{-0.73} \ \text{MeV}\\
    \Gamma^{\text{LC}}(X(4500) \rightarrow D_s D_s^*)= 2.02^{+0.26}_{-0.71} \ \text{MeV}\\
\end{eqnarray}

In order to obtain the coupling constant of $\hat{g}_{XJ/\psi \phi}^{TP}$ from the three-point sum rules,
we need to determine the continuum threshold $s_0$, the Borel mass $M^2$ and $Q^2=-q^2$ where $q^2$ represents the momentum of $\phi$.
The same value of $s_0$ can be used as we have obtained in the two-point sum rules,
so that Eq.(\ref{Eq:s0}) is applied.

\begin{figure}
\centering
  \includegraphics[width=7cm]{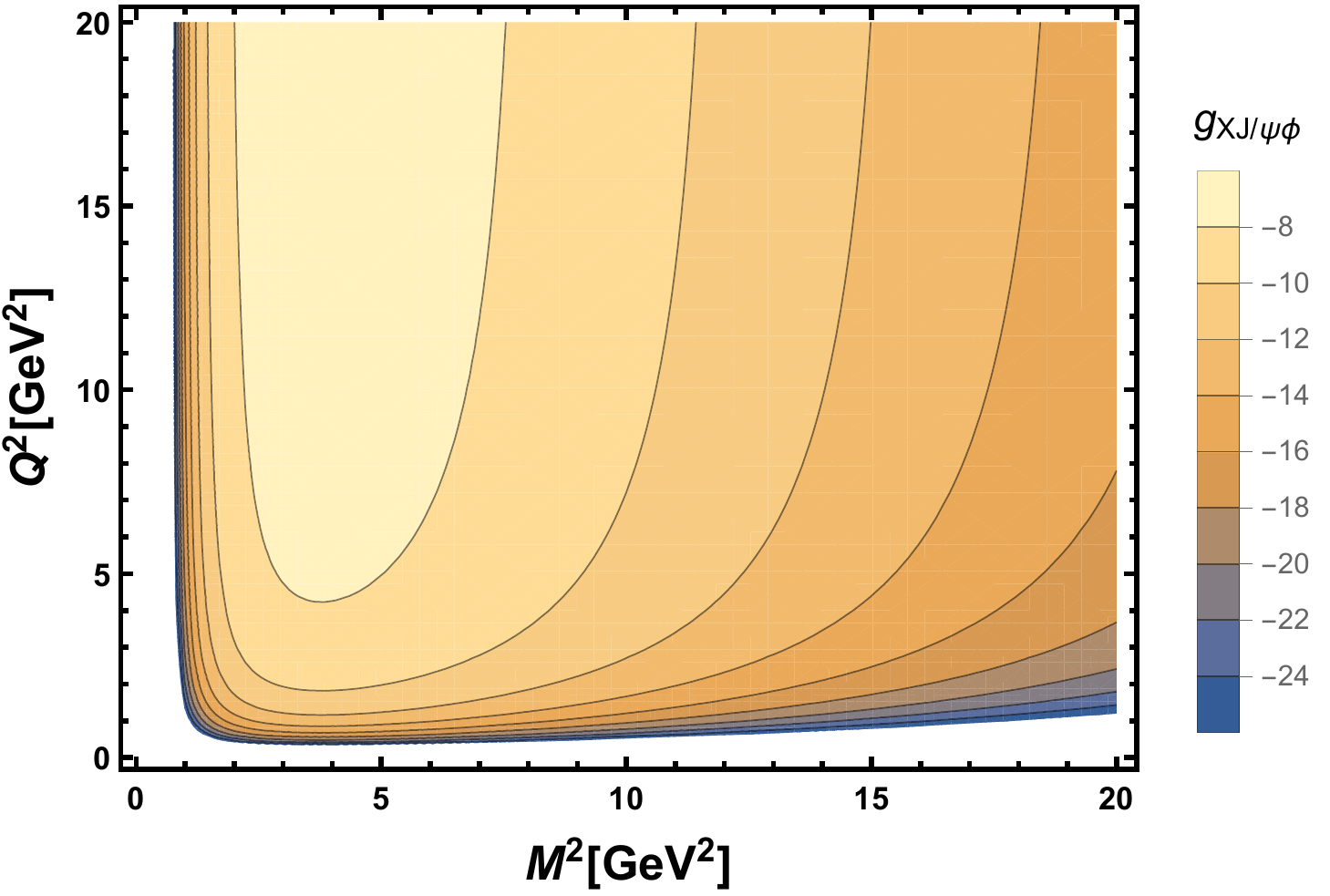}
  \caption{
The strong coupling $\hat{g}_{XJ/\psi \phi}^{TP}$ of a $X(4500)$ as a function of the $M^2$ and $Q^2$.}
\label{Fig:fig4a}
\end{figure}

As we see in Eq.\eqref{Three-Point}, the coupling constant $\hat{g}_{XJ/\psi \phi}^{TP}$ depends on two parameters, $Q^2$ and $M^2$.
An ideal working region can be determined by the fact that $\hat{g}_{XJ/\psi \phi}^{TP}$ should be independent as much as possible of $M^2$. 
In FIG.\ref{Fig:fig4a}, we show the  $\hat{g}_{XJ/\psi \phi}^{TP}$ as functions of $M^2$ in the contour map. 
As shown in the picture, the contour lines have Extreme Values of $M^2$ while 
\begin{eqnarray}
        3 \ \text{GeV}^2\leq M^2\leq 5 \ \text{GeV}^2
\end{eqnarray}
and
\begin{equation}
	\begin{aligned}
		Q^2\geq 8 \ \text{GeV}^2.
	\end{aligned}
\end{equation}

\begin{figure}[htbp]
\centering
  \includegraphics[width=6cm]{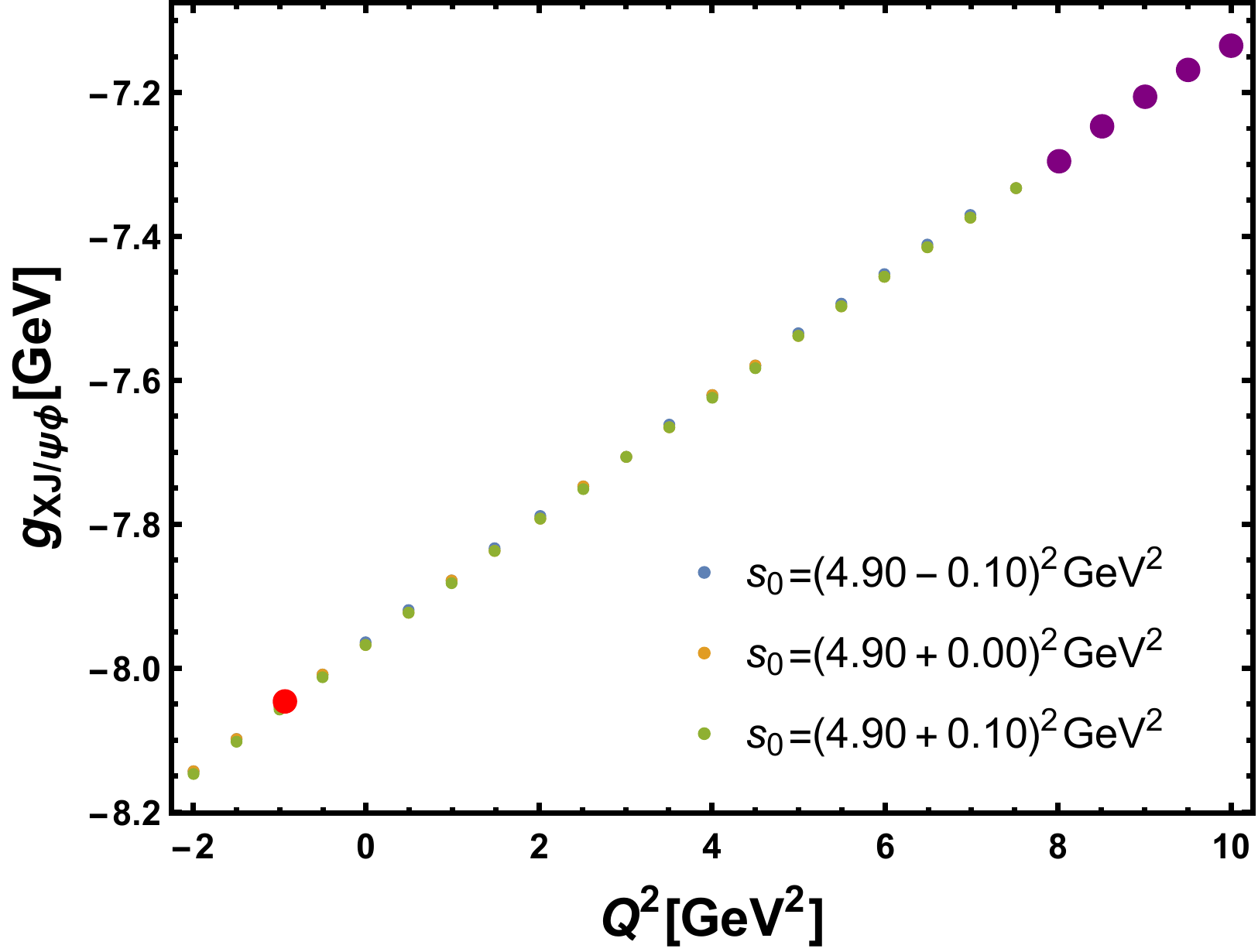}
  \caption{
  The exponential form of $\hat{g}_{XJ/\psi \phi}^{TP}$ as a function of the Borel Mass $Q^2$ at different fixed values of $s_0$ while $M^2=4.0$ GeV$^2$.}
\label{Fig:fig4b}
\end{figure}

Since the coupling constant is defined as the form factor at the pole of $Q^2=-m_\phi^2$, $\hat{g}_{XJ/\psi \phi}^{TP}$ should be extracted
from $Q^2=-m_\phi^2$ where the sum rules results become infinity and invalid.
However, our goal can be achieved by parametrizing the $\hat{g}_{XJ/\psi \phi}^{TP}$ with the help of \cite{Dias:2013xfa}:
\begin{eqnarray}\label{fit}
 		\hat{g}_{XJ/\psi \phi}^{TP}=A_1 e^{-\frac{A_2}{Q^2}}.
\end{eqnarray}
What we have to do is fitting the results of $\hat{g}_{XJ/\psi\phi}^{TP}$ in the region of $Q^2\geq 8\ \text{GeV}^2$ with the above equation.
Here, for convenient, we choose the region to be $8\ \text{GeV}^2 \leq Q^2 \leq 10\ \text{GeV}^2 $.
After those preparations, we can solve out $A_1$ and $A_2$ and receive the vaules that $A_1=7.9638$ and $A_2=90.698$. And we calculate the value $\hat{g}_{XJ/\psi \phi}^{TP}$
from $Q^2=-m_\phi^2$.
The exponential form of $\hat{g}_{XJ/\psi\phi}^{TP}$ is pictured in Fig.\ref{Fig:fig4b}. Fig.\ref{Fig:fig4b} shows the $Q^2$ dependence of $\hat{g}_{XJ/\psi \phi}^{TP}$ while we set $M^2=6.0$ $\rm GeV^2$.
For other values of $M^2$, the results are similar in the range of $5$ $\rm GeV^2$ $\leq M^2\leq 9$ $\rm GeV^2$. The purple dots are the data we tried to fit and the
red one shows the results at $Q^2=-m_\phi^2$:
\begin{eqnarray}
\hat{g}_{XJ/\psi \phi}^{TP}=-8.047^{+0.41}_{-0.42} \mathrm{GeV}.
\end{eqnarray}

\begin{figure}[htbp]
\centering
  \includegraphics[width=6cm]{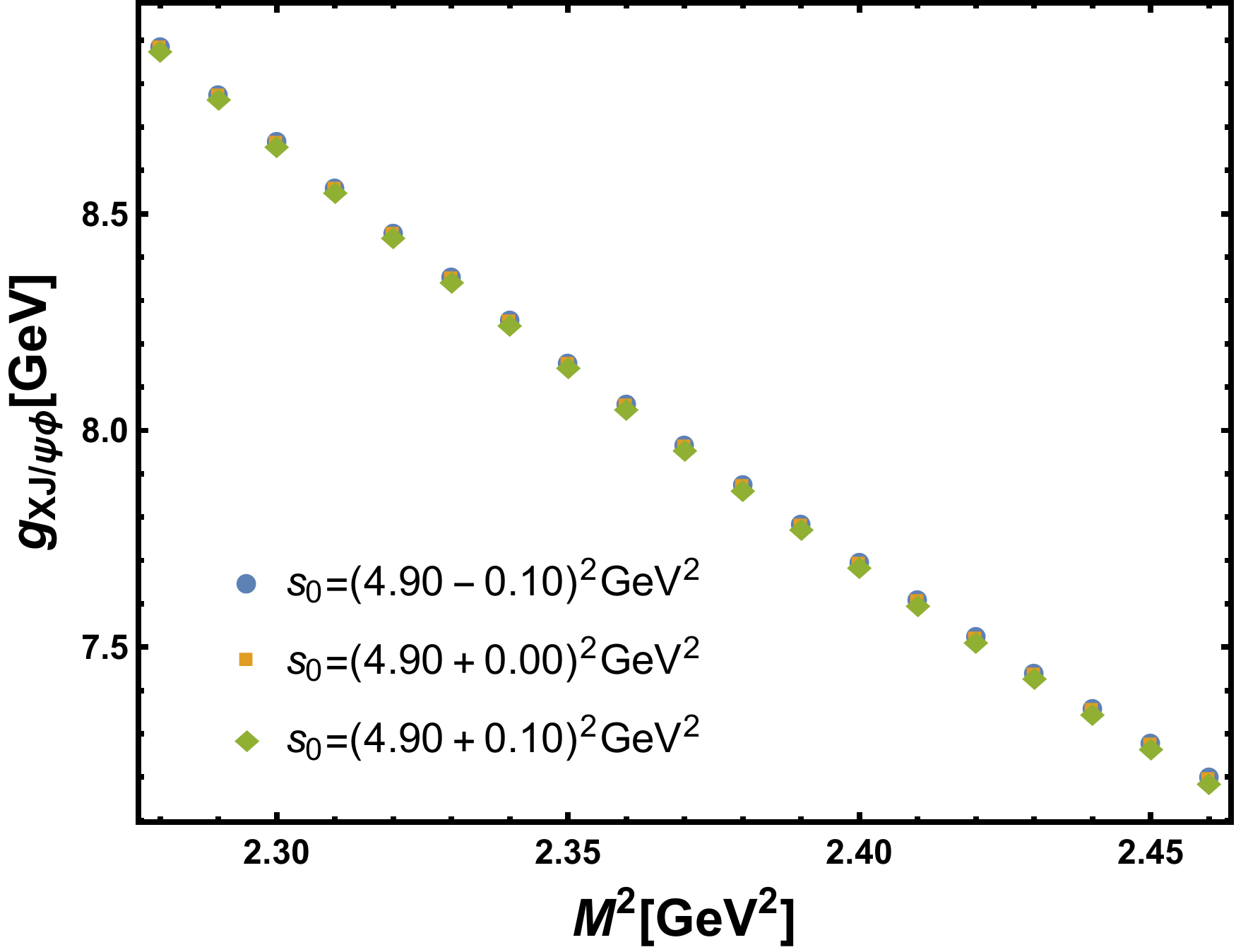}
  \caption{
 The strong coupling $ \tilde{g}_{XJ/\psi \phi}^{LC}$ in light-cone sum rules as a function of the Borel Mass $M^2$ at different fixed values of $s_0$.}
\label{Fig:fig5}
\end{figure}

Next, we pay attention to the coupling constant of $g_{XJ/\psi \phi}$ in the light-cone sum rules.
For $M^2$ and $s_0$, we use the same values as in the analysis of the mass. The prediction for $\tilde{g}_{XJ/\psi \phi}^{LC}$ is
\begin{equation}  
	\begin{aligned}
	\tilde{g}_{XJ/\psi \phi}^{LC}=7.796^{+1.191}_{-0.583}\ \text{GeV}.
	\end{aligned}
\end{equation}
By using the Eq.\eqref{couplingn}, the width of $X$(4500) $\to$ $J/\psi \phi$ are obtained to be
\begin{equation}
	\begin{aligned}
	\Gamma^{\text{TP}}(X(4500) \rightarrow J/\psi \phi)=124.32^{+12}_{-13}\ \text{MeV}
	\end{aligned}
\end{equation}
and 
\begin{equation}
	\begin{aligned}
	\Gamma^{\text{LC}}(X(4500) \rightarrow J/\psi \phi)= 116_{-38}^{+17} \ \text{MeV}
	\end{aligned}
\end{equation}
for the three-point sum rules and the light-cone sum rules respectively. Base on the experiment~\cite{LHCb:2021uow},
$X(4500)$ has total width of $77\pm6^{+10}_{-8}$ $\rm MeV$.
Both of the three-point sum rules prediction and that of the LCSR are close with the total width of $X$(4500) within the error.
From the calculation of $X(4500)\to D_sD_s$, $X(4500)\to D_s^*D_s$, and $X(4500)\to  D_s^*D_s^*$,
we can see the widths are much smaller than that of $X$(4500) $\to$ $J/\psi \phi$. The results of the open-charm decay channels, combined with the results of the hidden-charm decay channel, do not exceed the total width of the state. This suggest that the hidden-charm decay channels $X(4500)$ $\to$ $J/\psi\phi$ are predominant when compared with the total width of $X(4500)$ when we assign $X(4500)$ as a D-wave $cs\bar{c}\bar{s}$ tetraquark state. Moreover, calculating the complete open-charm decay will provide a more rational conclusion. If all those widths of open and hidden decays are consistent with the total width of $X(4500)$, the D-wave assignment for $X(4500)$ may be appropriate, and $X$(4500) $\to$ $J/\psi \phi$ will be its most significant decay channel. Fulture experiments are needed to further determine the structure of $X(4500)$.

\section{Summary}
\label{sec:summary}

By assigning a D-wave tetraquark state to $X$(4500), 
we investigate its mass, decay constant, and its decay of $X(4500)\to J/\psi\phi$ in this paper.
By evaluating the mass of $X(4500)$ via the two-point sum rules,
we found that the result was consistent with that in PDG. 
Meanwhile, the decay constant of $X(4500)$ has also been calculated. 
Then through both the approaches of the light-cone sum rules and the three-point sum rules, 
we calculate the strong coupling constant of $g_{X J/\psi\phi}$ and obtain the decay width.
In the case of D-wave tetraquarks, $X(4500)\to J/\psi\phi$ will be close to the total width of $X(4500)$.
Since our approach focuses exclusively on hidden-charm meson decay in this paper,
so we recommends calculating open-charm decays when $X(4500)$ is assigned as a state of D-wave tetraquarks.
D-wave interpretation is not appropriate for state $X(4500)$ when results from the open-charm decay channels plus those from the hidden decay channel exceed the width of the state.
Else we can conclude that D-wave tetraquark state may be appropriate for $X(4500)$
and that $X(4500)\to J/\psi\phi$ is the predominant process.
The results are instructive for future experiments to further determine the structure of $X(4500)$.


\begin{acknowledgments}
\noindent
Hao Sun is supported by the National Natural Science Foundation of China (Grant No.12075043, No.12147205).
\end{acknowledgments}


\section{Appendix}

\subsection{OPE results of open-charm decays}
\label{appendix:C}

\begin{eqnarray}
&&\Pi^{DD}_1(M^2)=\\\nonumber
&&-\frac{8 \pi ^8 m_c^2 m_s^2 \langle \bar{q}g_s\sigma Gq \rangle \langle \bar{q}q\rangle  e^{-\frac{m_c^2}{M^2}}}{M^4}+\frac{128 \pi ^9 \alpha_s m_c^2 m_s \langle \bar{q}q\rangle ^3 e^{-\frac{m_c^2}{M^2}}}{27 M^4}\\\nonumber
&&+\frac{\pi ^8 m_c^6 m_s^2 \langle \bar{q}g_s\sigma Gq \rangle^2 e^{-\frac{m_c^2}{M^2}}}{15 M^{10}}-\frac{16 \pi ^9 \alpha_s m_c^6 m_s \langle \bar{q}g_s\sigma Gq \rangle \langle \bar{q}q\rangle ^2 e^{-\frac{m_c^2}{M^2}}}{405 M^{10}}\\\nonumber
&&-\frac{\pi ^8 m_c^4 m_s^2 \langle \bar{q}g_s\sigma Gq \rangle^2 e^{-\frac{m_c^2}{M^2}}}{3 M^8}+\frac{16 \pi ^9 \alpha_s m_c^4 m_s \langle \bar{q}g_s\sigma Gq \rangle \langle \bar{q}q\rangle ^2 e^{-\frac{m_c^2}{M^2}}}{81 M^8}\\\nonumber
&&+\frac{8 \pi ^8 m_c^4 m_s^2 \langle \bar{q}g_s\sigma Gq \rangle \langle \bar{q}q\rangle  e^{-\frac{m_c^2}{M^2}}}{3 M^6}-\frac{128 \pi ^9 \alpha_s m_c^4 m_s \langle \bar{q}q\rangle ^3 e^{-\frac{m_c^2}{M^2}}}{81 M^6}
\end{eqnarray}

\begin{eqnarray}
&&\Pi^{DD}_2(M^2)= \int_{4m_c^2}^{s_0}e^{-s/M^2}ds\\\nonumber
&&\frac{32 m_s \pi ^8 \langle \bar{q}g_s\sigma Gq \rangle \langle \bar{q}q\rangle  s m_c^9}{\left(m^2-s\right)^7}-\frac{512 \pi ^9 \langle \bar{q}q\rangle ^3 s \alpha_s m_c^9}{27 \left(m_c^2-s\right)^7}\\\nonumber
&&+\frac{320 \langle g_s^2GG\rangle \pi ^{10} \langle \bar{q}q\rangle ^2 s m_c^8}{3 \left(m^2-s\right)^7}+\frac{12 m_s \pi ^8 \langle \bar{q}g_s\sigma Gq \rangle^2 s m_c^7}{\left(m_c^2-s\right)^7}\\\nonumber
&&+\frac{32 \langle g_s^2GG\rangle m_s \pi ^{10} \langle \bar{q}q\rangle ^2 s m_c^7}{3 \left(m_c^2-s\right)^7}+\frac{2048 \pi ^9 \langle \bar{q}q\rangle ^3 s^2 \alpha_s m_c^7}{27 \left(m_c^2-s\right)^7}\\\nonumber
&&-\frac{128 m_s \pi ^8 \langle \bar{q}g_s\sigma Gq \rangle \langle \bar{q}q\rangle  s^2 m_c^7}{\left(m_c^2-s\right)^7}-\frac{64 \pi ^9 \langle \bar{q}g_s\sigma Gq \rangle \langle \bar{q}q\rangle ^2 s \alpha_s m_c^7}{9 \left(m_c^2-s\right)^7}\\\nonumber
&&+\frac{40 \langle g_s^2GG\rangle \pi ^{10} \langle \bar{q}g_s\sigma Gq \rangle \langle \bar{q}q\rangle  s m_c^6}{\left(m_c^2-s\right)^7}\\ \nonumber
&&-\frac{3200 \langle g_s^2GG\rangle \pi ^{10} \langle \bar{q}q\rangle ^2 s^2 m_c^6}{9 \left(m_c^2-s\right)^7}\\\nonumber
&&+\frac{192 m_s \pi ^8 \langle \bar{q}g_s\sigma Gq \rangle \langle \bar{q}q\rangle  s^3 m_c^5}{\left(m_c^2-s\right)^7}\\ \nonumber
&&+\frac{16 \langle g_s^2GG\rangle m_s \pi ^{10} \langle \bar{q}g_s\sigma Gq \rangle \langle \bar{q}q\rangle  s m_c^5}{3 \left(m_c^2-s\right)^7}\\\nonumber
&&+\frac{448 \pi ^9 \langle \bar{q}g_s\sigma Gq \rangle \langle \bar{q}q\rangle ^2 s^2 \alpha_s m_c^5}{27 \left(m_c^2-s\right)^7}-\frac{28 m_s \pi ^8 \langle \bar{q}g_s\sigma Gq \rangle^2 s^2 m_c^5}{\left(m_c^2-s\right)^7}\\\nonumber
&&-\frac{224 \langle g_s^2GG\rangle m_s \pi ^{10} \langle \bar{q}q\rangle ^2 s^2 m_c^5}{9 \left(m_c^2-s\right)^7}-\frac{1024 \pi ^9 \langle \bar{q}q\rangle ^3 s^3 \alpha_s m_c^5}{9 \left(m_c^2-s\right)^7}\\\nonumber
&&+\frac{1280 \langle g_s^2GG\rangle \pi ^{10} \langle \bar{q}q\rangle ^2 s^3 m_c^4}{3 \left(m_c^2-s\right)^7}\\ \nonumber
&&-\frac{200 \langle g_s^2GG\rangle \pi ^{10} \langle \bar{q}g_s\sigma Gq \rangle \langle \bar{q}q\rangle  s^2 m_c^4}{3 \left(m_c^2-s\right)^7}\\\nonumber
&&+\frac{20 m_s \pi ^8 \langle \bar{q}g_s\sigma Gq \rangle^2 s^3 m_c^3}{\left(m_c^2-s\right)^7}+\frac{160 \langle g_s^2GG\rangle m_s \pi ^{10} \langle \bar{q}q\rangle ^2 s^3 m_c^3}{9 \left(m_c^2-s\right)^7}\\\nonumber
&&+\frac{2048 \pi ^9 \langle \bar{q}q\rangle ^3 s^4 \alpha_s m_c^3}{27 \left(m_c^2-s\right)^7}-\frac{128 m_s \pi ^8 \langle \bar{q}g_s\sigma Gq \rangle \langle \bar{q}q\rangle  s^4 m_c^3}{\left(m_c^2-s\right)^7}\\\nonumber
&&-\frac{16 \langle g_s^2GG\rangle m_s \pi ^{10} \langle \bar{q}g_s\sigma Gq \rangle \langle \bar{q}q\rangle  s^2 m_c^3}{9 \left(m_c^2-s\right)^7}\\ \nonumber
&&-\frac{320 \pi ^9 \langle \bar{q}g_s\sigma Gq \rangle \langle \bar{q}q\rangle ^2 s^3 \alpha_s m_c^3}{27 \left(m_c^2-s\right)^7}\\\nonumber
&&+\frac{280 \langle g_s^2GG\rangle \pi ^{10} \langle \bar{q}g_s\sigma Gq \rangle \langle \bar{q}q\rangle  s^3 m_c^2}{9 \left(m_c^2-s\right)^7}\\ \nonumber
&&-\frac{640 \langle g_s^2GG\rangle \pi ^{10} \langle \bar{q}q\rangle ^2 s^4 m_c^2}{3 \left(m_c^2-s\right)^7}\\\nonumber
&&+\frac{32 m_s \pi ^8 \langle \bar{q}g_s\sigma Gq \rangle \langle \bar{q}q\rangle  s^5 m_c}{\left(m_c^2-s\right)^7}+\frac{64 \pi ^9 \langle \bar{q}g_s\sigma Gq \rangle \langle \bar{q}q\rangle ^2 s^4 \alpha_s m_c}{27 \left(m_c^2-s\right)^7}\\\nonumber
&&-\frac{4 m_s \pi ^8 \langle \bar{q}g_s\sigma Gq \rangle^2 s^4 m_c}{\left(m_c^2-s\right)^7}-\frac{32 \langle g_s^2GG\rangle m_s \pi ^{10} \langle \bar{q}q\rangle ^2 s^4 m_c}{9 \left(m_c^2-s\right)^7}\\\nonumber
&&-\frac{512 \pi ^9 \langle \bar{q}q\rangle ^3 s^5 \alpha_s m_c}{27 \left(m_c^2-s\right)^7}+\frac{320 \langle g_s^2GG\rangle \pi ^{10} \langle \bar{q}q\rangle ^2 s^5}{9 \left(m_c^2-s\right)^7}\\\nonumber
&&-\frac{40 \langle g_s^2GG\rangle \pi ^{10} \langle \bar{q}g_s\sigma Gq \rangle \langle \bar{q}q\rangle  s^4}{9 \left(m_c^2-s\right)^7}
\end{eqnarray}

\begin{eqnarray}
&&\Pi^{DDs}_1(M^2)=\\\nonumber
&&\frac{20 \pi ^{10} \langle g_s^2GG\rangle m_c^2 \langle \bar{q}g_s\sigma Gq \rangle \langle \bar{q}q\rangle  e^{-\frac{m_c^2}{M^2}}}{9 M^6}\\ \nonumber
&&+\frac{16 \pi ^{10} \langle g_s^2GG\rangle m_c m_s \langle \bar{q}q\rangle ^2 e^{-\frac{m_c^2}{M^2}}}{9 M^4}\\\nonumber
&&+\frac{20 \pi ^{10} \langle g_s^2GG\rangle \langle \bar{q}g_s\sigma Gq \rangle \langle \bar{q}q\rangle  e^{-\frac{m_c^2}{M^2}}}{9 M^4}\\ \nonumber
&&+\frac{160 \pi ^{10} \langle g_s^2GG\rangle m_c^2 \langle \bar{q}q\rangle ^2 e^{-\frac{m_c^2}{M^2}}}{9 M^4}\\ \nonumber
&&+\frac{320 \pi ^{10} \langle g_s^2GG\rangle \langle \bar{q}q\rangle ^2 e^{-\frac{m_c^2}{M^2}}}{9 M^2}\\ \nonumber
&&-\frac{2 \pi ^{10} \langle g_s^2GG\rangle m_c^7 m_s \langle \bar{q}g_s\sigma Gq \rangle \langle \bar{q}q\rangle  e^{-\frac{m_c^2}{M^2}}}{405 M^12}\\ \nonumber
&&-\frac{4 \pi ^{10} \langle g_s^2GG\rangle m_c^6 \langle \bar{q}g_s\sigma Gq \rangle \langle \bar{q}q\rangle  e^{-\frac{m_c^2}{M^2}}}{27 M^10}\\ \nonumber
&&+\frac{2 \pi ^{10} \langle g_s^2GG\rangle m_c^5 m_s \langle \bar{q}g_s\sigma Gq \rangle \langle \bar{q}q\rangle  e^{-\frac{m_c^2}{M^2}}}{135 M^10}\\\nonumber
&&-\frac{8 \pi ^{10} \langle g_s^2GG\rangle m_c^5 m_s \langle \bar{q}q\rangle ^2 e^{-\frac{m_c^2}{M^2}}}{27 M^8}\\ \nonumber
&&-\frac{320 \pi ^{10} \langle g_s^2GG\rangle m_c^4 \langle \bar{q}q\rangle ^2 e^{-\frac{m_c^2}{M^2}}}{27 M^6}\\ \nonumber
&&+\frac{2 \pi ^{10} \langle g_s^2GG\rangle m_c^3 m_s \langle \bar{q}g_s\sigma Gq \rangle \langle \bar{q}q\rangle  e^{-\frac{m_c^2}{M^2}}}{27 M^8}\\ \nonumber
&&+\frac{16 \pi ^{10} \langle g_s^2GG\rangle m_c^3 m_s \langle \bar{q}q\rangle ^2 e^{-\frac{m_c^2}{M^2}}}{27 M^6}\\\nonumber
&&+\frac{2 \pi ^8 m_c m_s \langle \bar{q}g_s\sigma Gq \rangle^2 e^{-\frac{m_c^2}{M^2}}}{M^4}-\frac{32 \pi ^9 \alpha_s m_c \langle \bar{q}g_s\sigma Gq \rangle \langle \bar{q}q\rangle ^2 e^{-\frac{m_c^2}{M^2}}}{27 M^4}+\\ \nonumber
&&\frac{32 \pi ^8 m_c m_s \langle \bar{q}g_s\sigma Gq \rangle \langle \bar{q}q\rangle  e^{-\frac{m_c^2}{M^2}}}{M^2}-\frac{512 \pi ^9 \alpha_s m_c \langle \bar{q}q\rangle ^3 e^{-\frac{m_c^2}{M^2}}}{27 M^2}\\ \nonumber
&&-\frac{\pi ^8 m_c^5 m_s \langle \bar{q}g_s\sigma Gq \rangle^2 e^{-\frac{m_c^2}{M^2}}}{3 M^8}+\frac{16 \pi ^9 \alpha_s m_c^5 \langle \bar{q}g_s\sigma Gq \rangle \langle \bar{q}q\rangle ^2 e^{-\frac{m_c^2}{M^2}}}{81 M^8}\\ \nonumber
&&+\frac{2 \pi ^8 m_c^3 m_s \langle \bar{q}g_s\sigma Gq \rangle^2 e^{-\frac{m_c^2}{M^2}}}{3 M^6}-\frac{32 \pi ^9 \alpha_s m_c^3 \langle \bar{q}g_s\sigma Gq \rangle \langle \bar{q}q\rangle ^2 e^{-\frac{m_c^2}{M^2}}}{81 M^6}\\\nonumber
&&-\frac{16 \pi ^8 m_c^3 m_s \langle \bar{q}g_s\sigma Gq \rangle \langle \bar{q}q\rangle  e^{-\frac{m_c^2}{M^2}}}{M^4}+\frac{256 \pi ^9 \alpha_s m_c^3 \langle \bar{q}q\rangle ^3 e^{-\frac{m_c^2}{M^2}}}{27 M^4}
\end{eqnarray}
\begin{eqnarray}
\Pi^{DDs}_2(M^2)=0
\end{eqnarray}

\begin{eqnarray}
&&\Pi^{DsDs}_1(M^2)=\\\nonumber
&&\frac{64 \pi ^{10} \langle g_s^2GG\rangle m_c^2 \langle \bar{q}q\rangle ^2 e^{-\frac{m_c^2}{M^2}}}{27 M^4}\\ \nonumber
&&-\frac{8 \pi ^{10} \langle g_s^2GG\rangle m_c^8 \langle \bar{q}g_s\sigma Gq \rangle \langle \bar{q}q\rangle  e^{-\frac{m_c^2}{M^2}}}{2835 M^{14}}\\\nonumber
&&\frac{64 \pi ^{10} \langle g_s^2GG\rangle m_c^2 \langle \bar{q}q\rangle ^2 e^{-\frac{m_c^2}{M^2}}}{27 M^4}\\ \nonumber
&&-\frac{8 \pi ^{10} \langle g_s^2GG\rangle m_c^8 \langle \bar{q}g_s\sigma Gq \rangle \langle \bar{q}q\rangle  e^{-\frac{m_c^2}{M^2}}}{2835 M^{14}}\\\nonumber
&&+\frac{4 \pi ^{10} \langle g_s^2GG\rangle m_c^6 \langle \bar{q}g_s\sigma Gq \rangle \langle \bar{q}q\rangle  e^{-\frac{m_c^2}{M^2}}}{81 M^{12}}\\ \nonumber
&&-\frac{2 \pi ^{10} \langle g_s^2GG\rangle m_c^6 \langle \bar{q}g_s\sigma Gq \rangle \langle \bar{q}q\rangle  e^{-\frac{m_c^2}{M^2}}}{405 M^{10}}\\\nonumber
&&-\frac{16 \pi ^{10} \langle g_s^2GG\rangle m_c^4 \langle \bar{q}g_s\sigma Gq \rangle \langle \bar{q}q\rangle  e^{-\frac{m_c^2}{M^2}}}{135 M^{10}}\\ \nonumber
&&+\frac{4 \pi ^{10} \langle g_s^2GG\rangle m_c^4 \langle \bar{q}g_s\sigma Gq \rangle \langle \bar{q}q\rangle  e^{-\frac{m_c^2}{M^2}}}{81 M^8}\\ \nonumber
&&-\frac{32 \pi ^{10} \langle g_s^2GG\rangle m_c^4 \langle \bar{q}q\rangle ^2 e^{-\frac{m_c^2}{M^2}}}{81 M^6}\\ \nonumber
&&-\frac{16 \pi ^8 m_c^2 m_s^2 \langle \bar{q}g_s\sigma Gq \rangle \langle \bar{q}q\rangle  e^{-\frac{m^2}{M^2}}}{3 M^4}\\ \nonumber
&&+\frac{256 \pi ^9 \alpha_s m_c^2 m_s \langle \bar{q}q\rangle ^3 e^{-\frac{m_c^2}{M^2}}}{81 M^4}\\\nonumber
&&+\frac{8 \pi ^8 m_c^8 m_s^2 \langle \bar{q}g_s\sigma Gq \rangle^2 e^{-\frac{m_c^2}{M^2}}}{189 M^{14}}\\ \nonumber
&&-\frac{128 \pi ^9 \alpha_s m_c^8 m_s \langle \bar{q}g_s\sigma Gq \rangle \langle \bar{q}q\rangle ^2 e^{-\frac{m_c^2}{M^2}}}{5103 M^{14}}\\ \nonumber
&&-\frac{56 \pi ^8 m_c^6 m_s^2 \langle \bar{q}g_s\sigma Gq \rangle^2 e^{-\frac{m_c^2}{M^2}}}{135 M^{12}}\\ \nonumber
&&+\frac{896 \pi ^9 \alpha_s m_c^6 m_s \langle \bar{q}g_s\sigma Gq \rangle \langle \bar{q}q\rangle ^2 e^{-\frac{m_c^2}{M^2}}}{3645 M^12}\\ \nonumber
&&+\frac{2 \pi ^8 m_c^6 m_s^2 \langle \bar{q}g_s\sigma Gq \rangle^2 e^{-\frac{m_c^2}{M^2}}}{45 M^{10}}\\ \nonumber
&&-\frac{32 \pi ^9 \alpha_s m_c^6 m_s \langle \bar{q}g_s\sigma Gq \rangle \langle \bar{q}q\rangle ^2 e^{-\frac{m_c^2}{M^2}}}{1215 M^{10}}\\ \nonumber
&&+\frac{32 \pi ^8 m_c^4 m_s^2 \langle \bar{q}g_s\sigma Gq \rangle^2 e^{-\frac{m_c^2}{M^2}}}{45 M^{10}}\\ \nonumber
&&-\frac{512 \pi ^9 \alpha_s m_c^4 m_s \langle \bar{q}g_s\sigma Gq \rangle \langle \bar{q}q\rangle ^2 e^{-\frac{m_c^2}{M^2}}}{1215 M^{10}}\\ \nonumber
&&-\frac{2 \pi ^8 m_c^4 m_s^2 \langle \bar{q}g_s\sigma Gq \rangle^2 e^{-\frac{m_c^2}{M^2}}}{9 M^8}\\\nonumber
&&+\frac{32 \pi ^9 \alpha_s m_c^4 m_s \langle \bar{q}g_s\sigma Gq \rangle \langle \bar{q}q\rangle ^2 e^{-\frac{m_c^2}{M^2}}}{243 M^8}\\ \nonumber
&&+\frac{16 \pi ^8 m_c^4 m_s^2 \langle \bar{q}g_s\sigma Gq \rangle \langle \bar{q}q\rangle  e^{-\frac{m_c^2}{M^2}}}{9 M^6}\\ \nonumber
&&-\frac{256 \pi ^9 \alpha_s m_c^4 m_s \langle \bar{q}q\rangle ^3 e^{-\frac{m_c^2}{M^2}}}{243 M^6}
\end{eqnarray}

\begin{eqnarray}
&&\Pi^{DsDs}_2(M^2)=\int_{4m_c^2}^{u_0}e^{-s/M^2}ds\\\nonumber
&&\frac{64 m_s \pi ^8 \langle \bar{q}g_s\sigma Gq \rangle \langle \bar{q}q\rangle  m^{15}}{3 \left(m^2-s\right)^9}-\frac{1024 \pi ^9 \langle \bar{q}q\rangle ^3 \alpha_s m^{15}}{81 \left(m^2-s\right)^9}\\\nonumber
&&+\frac{256 m_s \pi ^9 \langle \bar{q}q\rangle ^3 \alpha_s m^{14}}{81 \left(m^2-s\right)^9}-\frac{16 m_s^2 \pi ^8 \langle \bar{q}g_s\sigma Gq \rangle \langle \bar{q}q\rangle  m^{14}}{3 \left(m^2-s\right)^9}\\\nonumber
&&+\frac{5888 \pi ^9 \langle \bar{q}q\rangle ^3 s \alpha_s m^{13}}{81 \left(m^2-s\right)^9}+\frac{16 m_s \pi ^8 \langle \bar{q}g_s\sigma Gq \rangle^2 m^{13}}{3 \left(m^2-s\right)^9}\\\nonumber
&&-\frac{368 m_s \pi ^8 \langle \bar{q}g_s\sigma Gq \rangle \langle \bar{q}q\rangle  s m^{13}}{3 \left(m^2-s\right)^9}-\frac{64 \langle g_s^2GG\rangle m_s \pi ^{10} \langle \bar{q}q\rangle ^2 m^{13}}{27 \left(m^2-s\right)^9}\\\nonumber
&&-\frac{256 \pi ^9 \langle \bar{q}g_s\sigma Gq \rangle \langle \bar{q}q\rangle ^2 \alpha_s m^{13}}{81 \left(m^2-s\right)^9}+\frac{32 m_s^2 \pi ^8 \langle \bar{q}g_s\sigma Gq \rangle \langle \bar{q}q\rangle  s m^{12}}{\left(m^2-s\right)^9}\\\nonumber
&&+\frac{448 \langle g_s^2GG\rangle \pi ^{10} \langle \bar{q}q\rangle ^2 s m^{12}}{9 \left(m^2-s\right)^9}+\frac{1024 m_s \pi ^9 \langle \bar{q}q\rangle ^3 \alpha_s m^{12}}{27 \left(m^2-s\right)^9}\\\nonumber
&&-\frac{64 m_s^2 \pi ^8 \langle \bar{q}g_s\sigma Gq \rangle \langle \bar{q}q\rangle  m^{12}}{\left(m^2-s\right)^9}+\frac{4 m_s^2 \pi ^8 \langle \bar{q}g_s\sigma Gq \rangle^2 m^{12}}{3 \left(m^2-s\right)^9}\\\nonumber
&&-\frac{512 m_s \pi ^9 \langle \bar{q}q\rangle ^3 s \alpha_s m^{12}}{27 \left(m^2-s\right)^9}-\frac{64 m_s \pi ^9 \langle \bar{q}g_s\sigma Gq \rangle \langle \bar{q}q\rangle ^2 \alpha_s m^{12}}{81 \left(m^2-s\right)^9}\\\nonumber
&&+\frac{288 m_s \pi ^8 \langle \bar{q}g_s\sigma Gq \rangle \langle \bar{q}q\rangle  s^2 m^{11}}{\left(m^2-s\right)^9}\\ \nonumber
&&+\frac{464 \langle g_s^2GG\rangle m_s \pi ^{10} \langle \bar{q}q\rangle ^2 s m^{11}}{27 \left(m^2-s\right)^9}\\\nonumber
&&+\frac{832 \pi ^9 \langle \bar{q}g_s\sigma Gq \rangle \langle \bar{q}q\rangle ^2 s \alpha_s m^{11}}{81 \left(m^2-s\right)^9}+\frac{64 m_s \pi ^8 \langle \bar{q}g_s\sigma Gq \rangle^2 m^{11}}{\left(m^2-s\right)^9}\\\nonumber
&&-\frac{52 m_s \pi ^8 \langle \bar{q}g_s\sigma Gq \rangle^2 s m^{11}}{3 \left(m^2-s\right)^9}-\frac{512 \pi ^9 \langle \bar{q}q\rangle ^3 s^2 \alpha_s m^{11}}{3 \left(m^2-s\right)^9}\\\nonumber
&&-\frac{64 \langle g_s^2GG\rangle m_s \pi ^{10} \langle \bar{q}q\rangle ^2 m^{11}}{9 \left(m^2-s\right)^9}\\ \nonumber
&&-\frac{8 \langle g_s^2GG\rangle m_s \pi ^{10} \langle \bar{q}g_s\sigma Gq \rangle \langle \bar{q}q\rangle  m^{11}}{27 \left(m^2-s\right)^9}\\\nonumber
&&-\frac{1024 \pi ^9 \langle \bar{q}g_s\sigma Gq \rangle \langle \bar{q}q\rangle ^2 \alpha_s m^{11}}{27 \left(m^2-s\right)^9}+\frac{240 m_s^2 \pi ^8 \langle \bar{q}g_s\sigma Gq \rangle \langle \bar{q}q\rangle  s m^{10}}{\left(m^2-s\right)^9}\\\nonumber
&&+\frac{56 \langle g_s^2GG\rangle \pi ^{10} \langle \bar{q}g_s\sigma Gq \rangle \langle \bar{q}q\rangle  s m^{10}}{3 \left(m^2-s\right)^9}+\frac{3904 m_s \pi ^9 \langle \bar{q}q\rangle ^3 s^2 \alpha_s m^{10}}{81 \left(m^2-s\right)^9}\\\nonumber
&&+\frac{928 m_s \pi ^9 \langle \bar{q}g_s\sigma Gq \rangle \langle \bar{q}q\rangle ^2 s \alpha_s m^{10}}{243 \left(m^2-s\right)^9}+\frac{64 m_s^2 \pi ^8 \langle \bar{q}g_s\sigma Gq \rangle^2 m^{10}}{3 \left(m^2-s\right)^9}\\\nonumber
&&-\frac{244 m_s^2 \pi ^8 \langle \bar{q}g_s\sigma Gq \rangle \langle \bar{q}q\rangle  s^2 m^{10}}{3 \left(m^2-s\right)^9}-\frac{58 m_s^2 \pi ^8 \langle \bar{q}g_s\sigma Gq \rangle^2 s m^{10}}{9 \left(m^2-s\right)^9}\\\nonumber
&&-\frac{1280 m_s \pi ^9 \langle \bar{q}q\rangle ^3 s \alpha_s m^{10}}{9 \left(m^2-s\right)^9}-\frac{7168 \langle g_s^2GG\rangle \pi ^{10} \langle \bar{q}q\rangle ^2 s^2 m^{10}}{27 \left(m^2-s\right)^9}\\\nonumber
&&-\frac{1024 m_s \pi ^9 \langle \bar{q}g_s\sigma Gq \rangle \langle \bar{q}q\rangle ^2 \alpha_s m^{10}}{81 \left(m^2-s\right)^9}+\frac{40 m_s \pi ^8 \langle \bar{q}g_s\sigma Gq \rangle^2 s^2 m^9}{3 \left(m^2-s\right)^9}\\\nonumber
&&+\frac{16 m_s \pi ^8 \langle \bar{q}g_s\sigma Gq \rangle^2 s m^9}{\left(m^2-s\right)^9}+\frac{256 \langle g_s^2GG\rangle m_s \pi ^{10} \langle \bar{q}q\rangle ^2 s m^9}{9 \left(m^2-s\right)^9}\\\nonumber
&&+\frac{224 \langle g_s^2GG\rangle m_s \pi ^{10} \langle \bar{q}g_s\sigma Gq \rangle \langle \bar{q}q\rangle  s m^9}{81 \left(m^2-s\right)^9}+\frac{16640 \pi ^9 \langle \bar{q}q\rangle ^3 s^3 \alpha_s m^9}{81 \left(m^2-s\right)^9}\\\nonumber
&&-\frac{1040 m_s \pi ^8 \langle \bar{q}g_s\sigma Gq \rangle \langle \bar{q}q\rangle  s^3 m^9}{3 \left(m^2-s\right)^9}-\frac{1280 \langle g_s^2GG\rangle m_s \pi ^{10} \langle \bar{q}q\rangle ^2 s^2 m^9}{27 \left(m^2-s\right)^9}\\\nonumber
&&-\frac{128 \langle g_s^2GG\rangle m_s \pi ^{10} \langle \bar{q}g_s\sigma Gq \rangle \langle \bar{q}q\rangle  m^9}{27 \left(m^2-s\right)^9}\\ \nonumber
&&-\frac{256 \pi ^9 \langle \bar{q}g_s\sigma Gq \rangle \langle \bar{q}q\rangle ^2 s \alpha_s m^9}{27 \left(m^2-s\right)^9}\\\nonumber
&&-\frac{640 \pi ^9 \langle \bar{q}g_s\sigma Gq \rangle \langle \bar{q}q\rangle ^2 s^2 \alpha_s m^9}{81 \left(m^2-s\right)^9}+\frac{340 m_s^2 \pi ^8 \langle \bar{q}g_s\sigma Gq \rangle \langle \bar{q}q\rangle  s^3 m^8}{3 \left(m^2-s\right)^9}\\\nonumber
&&+\frac{15680 \langle g_s^2GG\rangle \pi ^{10} \langle \bar{q}q\rangle ^2 s^3 m^8}{27 \left(m^2-s\right)^9}+\frac{38 m_s^2 \pi ^8 \langle \bar{q}g_s\sigma Gq \rangle^2 s^2 m^8}{3 \left(m^2-s\right)^9}\\\nonumber
&&+\frac{896 \langle g_s^2GG\rangle \pi ^{10} \langle \bar{q}g_s\sigma Gq \rangle \langle \bar{q}q\rangle  s m^8}{3 \left(m^2-s\right)^9}+\frac{1792 m_s \pi ^9 \langle \bar{q}q\rangle ^3 s^2 \alpha_s m^8}{9 \left(m^2-s\right)^9}\\\nonumber
&&+\frac{2560 m_s \pi ^9 \langle \bar{q}g_s\sigma Gq \rangle \langle \bar{q}q\rangle ^2 s \alpha_s m^8}{81 \left(m^2-s\right)^9}\\ \nonumber
&&-\frac{336 m_s^2 \pi ^8 \langle \bar{q}g_s\sigma Gq \rangle \langle \bar{q}q\rangle  s^2 m^8}{\left(m^2-s\right)^9}\\\nonumber
&&-\frac{160 m_s^2 \pi ^8 \langle \bar{q}g_s\sigma Gq \rangle^2 s m^8}{3 \left(m^2-s\right)^9}\\ \nonumber
&&-\frac{616 \langle g_s^2GG\rangle \pi ^{10} \langle \bar{q}g_s\sigma Gq \rangle \langle \bar{q}q\rangle  s^2 m^8}{9 \left(m^2-s\right)^9}\\\nonumber
&&-\frac{5440 m_s \pi ^9 \langle \bar{q}q\rangle ^3 s^3 \alpha_s m^8}{81 \left(m^2-s\right)^9}-\frac{608 m_s \pi ^9 \langle \bar{q}g_s\sigma Gq \rangle \langle \bar{q}q\rangle ^2 s^2 \alpha_s m^8}{81 \left(m^2-s\right)^9}\\\nonumber
&&+\frac{640 m_s \pi ^8 \langle \bar{q}g_s\sigma Gq \rangle \langle \bar{q}q\rangle  s^4 m^7}{3 \left(m^2-s\right)^9}+\frac{40 m_s \pi ^8 \langle \bar{q}g_s\sigma Gq \rangle^2 s^3 m^7}{3 \left(m^2-s\right)^9}\\\nonumber
&&+\frac{1760 \langle g_s^2GG\rangle m_s \pi ^{10} \langle \bar{q}q\rangle ^2 s^3 m^7}{27 \left(m^2-s\right)^9}\\ \nonumber
&&+\frac{1280 \langle g_s^2GG\rangle m_s \pi ^{10} \langle \bar{q}g_s\sigma Gq \rangle \langle \bar{q}q\rangle  s m^7}{27 \left(m^2-s\right)^9}\\\nonumber
&&+\frac{4352 \pi ^9 \langle \bar{q}g_s\sigma Gq \rangle \langle \bar{q}q\rangle ^2 s^2 \alpha_s m^7}{27 \left(m^2-s\right)^9}-\frac{272 m_s \pi ^8 \langle \bar{q}g_s\sigma Gq \rangle^2 s^2 m^7}{\left(m^2-s\right)^9}\\\nonumber
&&-\frac{16 \langle g_s^2GG\rangle m_s \pi ^{10} \langle \bar{q}g_s\sigma Gq \rangle \langle \bar{q}q\rangle  s^2 m^7}{3 \left(m^2-s\right)^9}\\ \nonumber
&&-\frac{320 \langle g_s^2GG\rangle m_s \pi ^{10} \langle \bar{q}q\rangle ^2 s^2 m^7}{9 \left(m^2-s\right)^9}\\\nonumber
&&-\frac{10240 \pi ^9 \langle \bar{q}q\rangle ^3 s^4 \alpha_s m^7}{81 \left(m^2-s\right)^9}-\frac{640 \pi ^9 \langle \bar{q}g_s\sigma Gq \rangle \langle \bar{q}q\rangle ^2 s^3 \alpha_s m^7}{81 \left(m^2-s\right)^9}\\\nonumber
&&+\frac{208 m_s^2 \pi ^8 \langle \bar{q}g_s\sigma Gq \rangle \langle \bar{q}q\rangle  s^3 m^6}{\left(m^2-s\right)^9}\\ \nonumber
&&+\frac{2576 \langle g_s^2GG\rangle \pi ^{10} \langle \bar{q}g_s\sigma Gq \rangle \langle \bar{q}q\rangle  s^3 m^6}{27 \left(m^2-s\right)^9}\\\nonumber
&&+\frac{128 m_s^2 \pi ^8 \langle \bar{q}g_s\sigma Gq \rangle^2 s^2 m^6}{3 \left(m^2-s\right)^9}+\frac{4480 m_s \pi ^9 \langle \bar{q}q\rangle ^3 s^4 \alpha_s m^6}{81 \left(m^2-s\right)^9}\\\nonumber
&&+\frac{1856 m_s \pi ^9 \langle \bar{q}g_s\sigma Gq \rangle \langle \bar{q}q\rangle ^2 s^3 \alpha_s m^6}{243 \left(m^2-s\right)^9}\\ \nonumber
&&-\frac{280 m_s^2 \pi ^8 \langle \bar{q}g_s\sigma Gq \rangle \langle \bar{q}q\rangle  s^4 m^6}{3 \left(m^2-s\right)^9}\\\nonumber
&&-\frac{116 m_s^2 \pi ^8 \langle \bar{q}g_s\sigma Gq \rangle^2 s^3 m^6}{9 \left(m^2-s\right)^9}\\ \nonumber
&&-\frac{3584 \langle g_s^2GG\rangle \pi ^{10} \langle \bar{q}g_s\sigma Gq \rangle \langle \bar{q}q\rangle  s^2 m^6}{9 \left(m^2-s\right)^9}\\\nonumber
&&-\frac{17920 \langle g_s^2GG\rangle \pi ^{10} \langle \bar{q}q\rangle ^2 s^4 m^6}{27 \left(m^2-s\right)^9}\\\nonumber
&&-\frac{3328 m_s \pi ^9 \langle \bar{q}q\rangle ^3 s^3 \alpha_s m^6}{27 \left(m^2-s\right)^9}-\frac{2048 m_s \pi ^9 \langle \bar{q}g_s\sigma Gq \rangle \langle \bar{q}q\rangle ^2 s^2 \alpha_s m^6}{81 \left(m^2-s\right)^9}\\\nonumber
&&+\frac{240 m_s \pi ^8 \langle \bar{q}g_s\sigma Gq \rangle^2 s^3 m^5}{\left(m^2-s\right)^9}+\frac{128 \langle g_s^2GG\rangle m_s \pi ^{10} \langle \bar{q}q\rangle ^2 s^3 m^5}{9 \left(m^2-s\right)^9}\\\nonumber
&&+\frac{32 \langle g_s^2GG\rangle m_s \pi ^{10} \langle \bar{q}g_s\sigma Gq \rangle \langle \bar{q}q\rangle  s^3 m^5}{9 \left(m^2-s\right)^9}\\ \nonumber
&&+\frac{128 \langle g_s^2GG\rangle m_s \pi ^{10} \langle \bar{q}g_s\sigma Gq \rangle \langle \bar{q}q\rangle  s^2 m^5}{9 \left(m^2-s\right)^9}\\\nonumber
&&+\frac{256 \pi ^9 \langle \bar{q}q\rangle ^3 s^5 \alpha_s m^5}{9 \left(m^2-s\right)^9}+\frac{1280 \pi ^9 \langle \bar{q}g_s\sigma Gq \rangle \langle \bar{q}q\rangle ^2 s^4 \alpha_s m^5}{81 \left(m^2-s\right)^9}\\\nonumber
&&-\frac{48 m_s \pi ^8 \langle \bar{q}g_s\sigma Gq \rangle \langle \bar{q}q\rangle  s^5 m^5}{\left(m^2-s\right)^9}-\frac{80 m_s \pi ^8 \langle \bar{q}g_s\sigma Gq \rangle^2 s^4 m^5}{3 \left(m^2-s\right)^9}\\\nonumber
&&-\frac{1280 \pi ^9 \langle \bar{q}g_s\sigma Gq \rangle \langle \bar{q}q\rangle ^2 s^3 \alpha_s m^5}{9 \left(m^2-s\right)^9}\\ \nonumber
&&-\frac{1280 \langle g_s^2GG\rangle m_s \pi ^{10} \langle \bar{q}q\rangle ^2 s^4 m^5}{27 \left(m^2-s\right)^9}\\\nonumber
&&+\frac{136 m_s^2 \pi ^8 \langle \bar{q}g_s\sigma Gq \rangle \langle \bar{q}q\rangle  s^5 m^4}{3 \left(m^2-s\right)^9}+\frac{11200 \langle g_s^2GG\rangle \pi ^{10} \langle \bar{q}q\rangle ^2 s^5 m^4}{27 \left(m^2-s\right)^9}\\\nonumber
&&+\frac{64 m_s^2 \pi ^8 \langle \bar{q}g_s\sigma Gq \rangle^2 s^4 m^4}{9 \left(m^2-s\right)^9}+\frac{896 \langle g_s^2GG\rangle \pi ^{10} \langle \bar{q}g_s\sigma Gq \rangle \langle \bar{q}q\rangle  s^3 m^4}{9 \left(m^2-s\right)^9}\\\nonumber
&&+\frac{256 m_s \pi ^9 \langle \bar{q}q\rangle ^3 s^4 \alpha_s m^4}{9 \left(m^2-s\right)^9}+\frac{512 m_s \pi ^9 \langle \bar{q}g_s\sigma Gq \rangle \langle \bar{q}q\rangle ^2 s^3 \alpha_s m^4}{81 \left(m^2-s\right)^9}\\\nonumber
&&-\frac{48 m_s^2 \pi ^8 \langle \bar{q}g_s\sigma Gq \rangle \langle \bar{q}q\rangle  s^4 m^4}{\left(m^2-s\right)^9}-\frac{32 m_s^2 \pi ^8 \langle \bar{q}g_s\sigma Gq \rangle^2 s^3 m^4}{3 \left(m^2-s\right)^9}\\\nonumber
&&-\frac{560 \langle g_s^2GG\rangle \pi ^{10} \langle \bar{q}g_s\sigma Gq \rangle \langle \bar{q}q\rangle  s^4 m^4}{9 \left(m^2-s\right)^9}-\frac{2176 m_s \pi ^9 \langle \bar{q}q\rangle ^3 s^5 \alpha_s m^4}{81 \left(m^2-s\right)^9}\\\nonumber
&&-\frac{1024 m_s \pi ^9 \langle \bar{q}g_s\sigma Gq \rangle \langle \bar{q}q\rangle ^2 s^4 \alpha_s m^4}{243 \left(m^2-s\right)^9}+\frac{44 m_s \pi ^8 \langle \bar{q}g_s\sigma Gq \rangle^2 s^5 m^3}{3 \left(m^2-s\right)^9}\\\nonumber
&&+\frac{464 \langle g_s^2GG\rangle m_s \pi ^{10} \langle \bar{q}q\rangle ^2 s^5 m^3}{27 \left(m^2-s\right)^9}+\frac{512 \pi ^9 \langle \bar{q}q\rangle ^3 s^6 \alpha_s m^3}{81 \left(m^2-s\right)^9}\\\nonumber
&&+\frac{256 \pi ^9 \langle \bar{q}g_s\sigma Gq \rangle \langle \bar{q}q\rangle ^2 s^4 \alpha_s m^3}{9 \left(m^2-s\right)^9}-\frac{48 m_s \pi ^8 \langle \bar{q}g_s\sigma Gq \rangle^2 s^4 m^3}{\left(m^2-s\right)^9}\\\nonumber
&&-\frac{32 m_s \pi ^8 \langle \bar{q}g_s\sigma Gq \rangle \langle \bar{q}q\rangle  s^6 m^3}{3 \left(m^2-s\right)^9}\\ \nonumber
&&-\frac{256 \langle g_s^2GG\rangle m_s \pi ^{10} \langle \bar{q}g_s\sigma Gq \rangle \langle \bar{q}q\rangle  s^3 m^3}{27 \left(m^2-s\right)^9}\\\nonumber
&&-\frac{56 \langle g_s^2GG\rangle m_s \pi ^{10} \langle \bar{q}g_s\sigma Gq \rangle \langle \bar{q}q\rangle  s^4 m^3}{81 \left(m^2-s\right)^9}\\ \nonumber
&&-\frac{704 \pi ^9 \langle \bar{q}g_s\sigma Gq \rangle \langle \bar{q}q\rangle ^2 s^5 \alpha_s m^3}{81 \left(m^2-s\right)^9}\\\nonumber
&&+\frac{56 \langle g_s^2GG\rangle \pi ^{10} \langle \bar{q}g_s\sigma Gq \rangle \langle \bar{q}q\rangle  s^5 m^2}{3 \left(m^2-s\right)^9}+\frac{64 m_s \pi ^9 \langle \bar{q}q\rangle ^3 s^6 \alpha_s m^2}{9 \left(m^2-s\right)^9}\\\nonumber
&&+\frac{32 m_s \pi ^9 \langle \bar{q}g_s\sigma Gq \rangle \langle \bar{q}q\rangle ^2 s^5 \alpha_s m^2}{27 \left(m^2-s\right)^9}-\frac{12 m_s^2 \pi ^8 \langle \bar{q}g_s\sigma Gq \rangle \langle \bar{q}q\rangle  s^6 m^2}{\left(m^2-s\right)^9}\\\nonumber
&&-\frac{2 m_s^2 \pi ^8 \langle \bar{q}g_s\sigma Gq \rangle^2 s^5 m^2}{\left(m^2-s\right)^9}-\frac{3584 \langle g_s^2GG\rangle \pi ^{10} \langle \bar{q}q\rangle ^2 s^6 m^2}{27 \left(m^2-s\right)^9}\\\nonumber
&&+\frac{16 m_s \pi ^8 \langle \bar{q}g_s\sigma Gq \rangle \langle \bar{q}q\rangle  s^7 m}{3 \left(m^2-s\right)^9}+\frac{128 \pi ^9 \langle \bar{q}g_s\sigma Gq \rangle \langle \bar{q}q\rangle ^2 s^6 \alpha_s m}{81 \left(m^2-s\right)^9}\\\nonumber
&&-\frac{8 m_s \pi ^8 \langle \bar{q}g_s\sigma Gq \rangle^2 s^6 m}{3 \left(m^2-s\right)^9}-\frac{64 \langle g_s^2GG\rangle m_s \pi ^{10} \langle \bar{q}q\rangle ^2 s^6 m}{27 \left(m^2-s\right)^9}\\\nonumber
&&-\frac{256 \pi ^9 \langle \bar{q}q\rangle ^3 s^7 \alpha_s m}{81 \left(m^2-s\right)^9}+\frac{4 m_s^2 \pi ^8 \langle \bar{q}g_s\sigma Gq \rangle \langle \bar{q}q\rangle  s^7}{3 \left(m^2-s\right)^9}\\\nonumber
&&+\frac{448 \langle g_s^2GG\rangle \pi ^{10} \langle \bar{q}q\rangle ^2 s^7}{27 \left(m^2-s\right)^9}+\frac{2 m_s^2 \pi ^8 \langle \bar{q}g_s\sigma Gq \rangle^2 s^6}{9 \left(m^2-s\right)^9}\\
&&-\frac{56 \langle g_s^2GG\rangle \pi ^{10} \langle \bar{q}g_s\sigma Gq \rangle \langle \bar{q}q\rangle  s^6}{27 \left(m^2-s\right)^9}-\frac{64 m_s \pi ^9 \langle \bar{q}q\rangle ^3 s^7 \alpha_s}{81 \left(m^2-s\right)^9}\\\nonumber
&&-\frac{32 m_s \pi ^9 \langle \bar{q}g_s\sigma Gq \rangle \langle \bar{q}q\rangle ^2 s^6 \alpha_s}{243 \left(m^2-s\right)^9}
\end{eqnarray}

\subsection{The relations between the light-cone distribution amplitudes (LCDAs) and the matrix elements}
\label{appendix:A}

The matrix elements of $\phi$ can be represented in the format of distributions amplitudes~\cite{Ball:1996tb,Ball:2007zt} as:
\begin{eqnarray}
&&  \left\langle \phi(P, \lambda)\left|\bar{q}(x) \gamma_{\rho} q(0)\right| 0\right\rangle \\ \nonumber
&=& P_{\rho}\left(e^{(\lambda)} x\right) f_{\phi} m_{\phi} \times \int_{0}^{1} d u e^{i u p x} \Phi_{\|}(u, \mu) \\ \nonumber
&+& e_{\rho}^{(\lambda)} f_{\phi} m_{\phi} \int_{0}^{1} d u e^{i u p x} g_{\perp}^{(v)}(u, \mu),
\end{eqnarray}
\begin{eqnarray} 
&& \left\langle \phi(P, \lambda)\left|\bar{q}(x) \gamma_{\theta} \gamma_{5} q(0)\right| 0\right\rangle \\ \nonumber
&=&-\frac{1}{4} \epsilon_{\theta \nu \rho \sigma} e^{(\lambda) \nu} P^{\rho} x^{\sigma} f_{\phi} m_{\phi} \times \int_{0}^{1} d u e^{i u p x} g_{\perp}^{(a)}(u, \mu),
\end{eqnarray}
\begin{eqnarray}
&& \left\langle \phi(P, \lambda)\left|\bar{q}(x) \sigma_{\rho \nu} q(0)\right| 0\right\rangle \\ \nonumber
&=&-i\left(e_{\rho}^{(\lambda)} P_{\nu}-e_{\nu}^{(\lambda)} P_{\rho}\right) \times f_{\phi}^{\perp} \int_{0}^{1} d u e^{i u p x} \phi_{2}^\perp(u,\mu), \quad ~~
\end{eqnarray}
where $P, e$ are the momentum and polarization vector of $\phi$ respectively. The superscript $\lambda$ denotes the polarisation of
the $\phi$ meson: $\lambda=\|(\perp)$ for longitudinal (transverse) polarisation. $f^\perp_\phi$ is ransversely
polarized $\phi$ meson decay constant.
\begin{eqnarray} 
&& \Phi_{\|}(u, \mu) = \\ \nonumber
&& \frac{1}{2} \left[\bar{u} \int_{0}^{u} d y \frac{\phi^{\|}_2(y, \mu)}{\bar{y}}-u \int_{u}^{1} d y \frac{\phi^{\|}_2(y, \mu)}{y}\right],
\end{eqnarray}
and
\begin{eqnarray} 
&& g_{\perp}^{(v), \text { twist } 2}(u, \mu) = \\ \nonumber
&& \frac{1}{2} \left[\int_{0}^{u} d y \frac{\phi^{\|}_2(y, \mu)}{\bar{y}} + \int_{u}^{1} d y \frac{\phi^{\|}_2(y, \mu)}{y}\right], \\
&& g_{\perp}^{(a), \text { twist } 2}(u, \mu) = \\ \nonumber
&& 2\left[\bar{u} \int_{0}^{u} d y \frac{\phi^{\|}_2(y, \mu)}{\bar{y}}+u \int_{u}^{1} d y \frac{\phi^{\|}_2(y, \mu)}{y}\right],
\end{eqnarray}
$\bar{u}$ and $\bar{y}$ represent $1-u$ and $1-y$ respectively.
By using the Gegenbauer polynomials, $\phi_{2}^{\|, \perp}$ is given as~\cite{Ball:2007zt}:
\begin{equation}
\phi_{2}^{\|, \perp}(u, \mu)=6 u \bar{u}\left\{1+\sum_{n=1}^{\infty} a_{n}^{\|, \perp}(\mu) C_{n}^{3 / 2}(2 u-1)\right\}.
\end{equation}
The coefficients of $a_n$ and $C_{n}^{3 / 2}(x)$, and more details are shown in Ref.\cite{Ball:1996tb}. 
Sum rules calculations are usually based on the Fock-Schwinger gauge:
\begin{equation}
	\begin{aligned}
		x^\mu A_\mu^a(x)=0.
	\end{aligned}
\end{equation}
Based on some deductions, we can derive these relations \cite{Gubler:2013moa}:
\begin{equation}
\begin{aligned}
A_{\mu}^{a}(x)=& \frac{1}{2} x^{v} G_{v \mu}^{a}(0)+\frac{1}{3} x^{v} x^{\alpha}\left[D_{\alpha} G_{v \mu}(0)\right]^{a} \\
&+\frac{1}{8} x^{v} x^{\alpha} x^{\beta}\left[D_{\alpha} D_{\beta} G_{v \mu}(0)\right]^{a}+\cdots.
\end{aligned}
\end{equation}
As a result, if we reintroduce the gluonic fields into the covariant derivative 
and compute $\braket{\phi(P)|\bar{q}(0)\overleftarrow{D}_\mu\overleftarrow{D}_\nu\Gamma^aq(0)|0}$, we get the following results:
\begin{eqnarray} \nonumber
		&&\braket{\phi(P)|\bar{q}(0)\overleftarrow{D}_\mu\overleftarrow{D}_\nu\Gamma^aq(0)|0}\\ \nonumber
		&&=\braket{\phi(P)|\partial_\mu\partial_\nu[\bar{q}(x)]|_{x=0}\Gamma^aq(0)|0}\\ \nonumber
		&&+\braket{\phi(P)|\bar{q}(0)\frac{igT^b}{2}G^b_{\mu\nu}\Gamma^aq(0)|0}\\ 
		&&=\partial_\mu\partial_\nu[\braket{\phi(P)|\bar{q}(x)\Gamma^aq(0)|0}]|_{x=0}
\end{eqnarray}
where the $\braket{\phi(P)|\bar{q}(x)\Gamma^a q(0)|0}$ type matrix has been provided previously.

\subsection{Spectral densities}
\label{appendix:B}

In this section we provide the spectral densities for $J^{X}$.
In the following expressions, $\mathrm{H}(x)$ is defined as:
\begin{eqnarray}
    \mathrm{H}(x)=
    \begin{cases}
    0&x\ge0\\
    1&x< 0 .
    \end{cases}
\end{eqnarray}
The spectral density for $J^{X}$ can be divide into:
\begin{eqnarray} \nonumber
\Tilde{\rho}^{OPE}(\hat{s}) &=&
\Tilde{\rho}^{\braket{\frac{\alpha_s}{\pi}GG}}(\hat{s})+\Tilde{\rho}^{\braket{\bar{q}g_s\sigma Gq}}(\hat{s}) \\ \nonumber
&+&\tilde{\rho}^{\braket{\frac{\alpha_s}{\pi}GG} \braket{\bar{q}q}}(\hat{s}) +\tilde{\rho}^{\braket{\bar{q}g_s\sigma Gq} \braket{\bar{q}q}}(\hat{s}) \\
&+&\tilde{\rho}^{\braket{\bar{q}g_s\sigma Gq} \braket{\frac{\alpha_s}{\pi}GG}}(\hat{s})+\Tilde{\rho}^{\braket{\bar{q}q}^2\braket{\frac{\alpha_s}{\pi}GG}}(\hat{s})~~~~
\end{eqnarray}
with
\begin{eqnarray} \nonumber
&&\Tilde{\rho}^{\braket{\frac{\alpha_s}{\pi}GG}}(\hat{s})=
\int_0^1dx\int_0^1dy\frac{-y^2\braket{\frac{\alpha_s}{\pi}GG}}{36864 \pi ^4 (x-1)^4 x^3 (y-1)^3}\\ \nonumber
&&\left(m_c^8 \left(2 x \left(18 y^4+54 y^3+49 y^2+33 y-3\right)\right.\right.\\ \nonumber
&+& \left.36 y^4+117 y^3+109 y^2+34 y+6\right)\\ \nonumber
&-& 12 m_c^6 \hat{s} x (y-1)x^2\left( x^2 \left(30 y^4+84 y^3+79 y^2+38 y-3\right)\right.\\ \nonumber
&+&\left.2 x \left(3 y^3+2 y^2-7 y+2\right)-30 y^4-90 y^3-83 y^2-24 y-1\right)\\ \nonumber
&+&12 m_c^4 \hat{s}^2 (x-1)^2 x^2 (y-1)^2 \left(x \left(90 y^4+240 y^3+224 y^2\right.\right.\\ \nonumber
&+&\left.\left.90 y-6\right)+\left(90 y^3+255 y^2+229 y+64\right) y\right)\\ \nonumber
&-&4 m_c^2 \hat{s}^3 (x-1)^3 x^3 (y-1)^3 \left(x \left(315 y^4+810 y^3+740 y^2\right.\right.\\ \nonumber
&+&\left.\left.264 y-15\right)+315 y^4+855 y^3+745 y^2+202 y-3\right)\\ \nonumber
&+&3 \hat{s}^4 (x-1)^4 x^4 (y-1)^4 \left(2 x \left(84 y^4+210 y^3+187 y^2+61 y\right.\right.\\ \nonumber
&-&\left.\left.\left.3\right)+168 y^4+441 y^3+373 y^2+98 y-2\right)\right)\\
&&\times H\left(y \left(x \hat{s}^2 (x (-y)+x+y-1)+m_c^2\right)\right),
\end{eqnarray}

\begin{eqnarray} \nonumber
&&\Tilde{\rho}^{\braket{\bar{q}g_s\sigma Gq}}(\hat{s})=
\int_0^1dx\int_0^1dy\frac{-m_s\braket{\bar{q}g_s\sigma Gq}}{192 \pi ^4 (x-1)^2 x^2 (y-1)^2}\\ \nonumber
&&\times\left( 2 m-c^6 y \left(6 y^3+7 y^2+8 y-1\right)\right.\\ \nonumber
&-&m_c^4 (x-1) x (y-1) \left(3 s y \left(24 y^3+33 y^2+26 y-3\right)\right.\\ \nonumber
&-&\left.m_0 \left(12 y^3+15 y^2+14 y-1\right)\right)\\ \nonumber
&+& m_c^2 \hat{s} (x-1)^2 x^2 (y-1)^2 \left(12 \hat{s} y \left(10 y^3+15 y^2+10 y-1\right)\right.\\ \nonumber
&-&\left.m_0 \left(48 y^3+69 y^2+50 y-3\right)\right)\\ \nonumber
&+& \hat{s}^2 (x-1)^3 x^3 (y-1)^3 \left(m_0 \left(40 y^3+62 y^2+40 y-2\right)\right.\\ \nonumber
&-&\left.\left.\hat{s} y \left(60 y^3+95 y^2+58 y-5\right)\right)\right.\\ 
&&\times H\left(y \left(x \hat{s}^2 (x (-y)+x+y-1)+m_c^2\right)\right),
\end{eqnarray}

\begin{eqnarray} \nonumber
&&\tilde{\rho}^{\braket{\frac{\alpha_s}{\pi}GG} \braket{\bar{q}q}}(\hat{s})=\int_0^1dx\int_0^1dy\frac{m_c  \braket{\frac{\alpha_s}{\pi}GG}\braket{\bar{q}q}}{32 \pi ^2 (x-1)^2 x (y-1)} \\ \nonumber
&&\times \left(m_c^2+\hat{s} x (x (-y)+x+y-1)m^2 (y+1) y\right.\\ \nonumber
&+&\left.2 m_c m_s (x-1) (y-1)-2 \hat{s} (x-1) x y \left(y^2-1\right)\right)\\
&&\times H\left(y \left(x \hat{s}^2 (x (-y)+x+y-1)+m_c^2\right)\right),
\end{eqnarray}

\begin{eqnarray} \nonumber
&&\tilde{\rho}^{\braket{\bar{q}g_s\sigma Gq} \braket{\bar{q}q}}(\hat{s})=\int_0^1dx\int_0^1dy\frac{g_s^2\braket{\bar{q}g_s\sigma Gq}\braket{\bar{q}q}}{1296 \pi ^4 (x-1) x (y-1)} \\ \nonumber
&&\times \left(m_c^4 \left(12 y^3+15 y^2+14 y-1\right)\right.\\ \nonumber
&-&\left.m-c^2 \hat{s} (x-1) x \left(48 y^4+21 y^3-19 y^2-53 y+3\right)\right.\\ \nonumber
&+&\left.2 \hat{s}^2 (x-1)^2 x^2 (y-1)^2 \left(20 y^3+31 y^2+20 y-1\right)\right)\\ 
&&\times H\left(y \left(x \hat{s}^2 (x (-y)+x+y-1)+m_c^2\right)\right),
\end{eqnarray}

\begin{eqnarray} \nonumber
&&\tilde{\rho}^{\braket{\bar{q}g_s\sigma Gq} \braket{\frac{\alpha_s}{\pi}GG}}(\hat{s})=\int_0^1dx\int_0^1dy\frac{m_c\braket{\bar{q}g_s\sigma Gq}\braket{\frac{\alpha_s}{\pi}GG}}{1296 \pi ^4 (x-1) x (y-1)} \\ \nonumber\\ \nonumber
&&\times (y+1) \left(2 m_c^2+3 \hat{s} x (x (-y)+x+y-1)\right)\\
&&\times H\left(y \left(x \hat{s}^2 (x (-y)+x+y-1)+m_c^2\right)\right),
\end{eqnarray}

\begin{eqnarray} \nonumber
&&\Tilde{\rho}^{\braket{\bar{q}q}^2\braket{\frac{\alpha_s}{\pi}GG}}(\hat{s})=\int_0^1dx \mathrm{H}\left(m_c^2-x (x+1) \hat{s}\right)\\
&&\times\frac{1}{24} \braket{\frac{\alpha_s}{\pi}GG} m_c \braket{\bar{q}q}^2\left(m_s x-2m_c\right).
\end{eqnarray}

\bibliography{ref}
\bibliographystyle{h-physrev5}
\end{document}